\documentclass[fleqn,usenatbib]{mnras}

\usepackage{newtxtext,newtxmath}

\usepackage[T1]{fontenc}
\usepackage{ae,aecompl}

\usepackage{graphicx}	
\usepackage{amsmath}	
\usepackage{pdflscape}

\title[Chemical abundances in PN M 2-36.]{Physical conditions and chemical abundances in PN M 2-36. Results from deep echelle observations}

\author[Esp\'iritu \& Peimbert]{
Jos\'e N. Esp\'iritu$^{1}$\thanks{E-mail: jespiritu@astro.unam.mx} \&
Antonio Peimbert$^{1}$
\\
$^{1}$Instituto de Astronom\'ia, Universidad Nacional Aut\'onoma de M\'exico, Apdo. Postal 70264, Ciudad de M\'exico 04510, M\'exico\\
}

\date{Accepted XXX. Received YYY; in original form ZZZ}

\pubyear{2021}

\begin{document}
\label{firstpage}
\pagerange{\pageref{firstpage}--\pageref{lastpage}}
\maketitle
	
\begin{abstract}
We present a spectrum of the planetary nebula M 2-36 obtained using the Ultra Violet and Visual Echelle Spectrograph (UVES) at the Very Large Telescope (VLT). 446 emission lines are detected. We perform an analysis of the chemical composition using multiple electron temperature ($T_{e}$) and density ($n_{e}$) diagnostics. $T_{e}$ and $n_{e}$ are computed using a variety of methods, including collisionally excited line (CEL) ratios, O$^{++}$ optical recombination lines (ORLs), and measuring the intensity of the Balmer jump. Besides the classical CEL abundances, we also present robust ionic abundances from ORLs of heavy elements. From CELs and ORLs of O$^{++}$, we obtain a new value for the Abundance Discrepancy Factor (ADF) of this nebula, being ADF(O$^{++})=$ 6.76 $\pm$ 0.50. From all the different line ratios that we study, we find that the object cannot be chemically homogeneous; moreover, we find that two-phased photoionization models are unable to simultaneously reproduce critical \ion{O}{ii} and [\ion{O}{iii}] line ratios. However, we find a three-phased model able to adequately reproduce such ratios. While we consider this to be a toy model, it is able to reproduce the observed temperature and density line diagnostics. Our analysis shows that it is important to study high ADF PNe with high spectral resolution, since its physical and chemical structure may be more complicated than previously thought.
\end{abstract}
	
	
	
	
    \section{Introduction}
    The study of chemical abundances in planetary nebulae (PNe) and \ion{H}{ii} regions has been mostly based on collisionally excited lines (CELs) of heavy elements, with a sizeable\footnote{Roger Wesson has compiled a list of objects with measured ADFs: https://www.nebulousresearch.org/adfs/} number of objects where optical recombination lines (ORLs) can also be used \citep[see for example][and references therein]{peimbert17}. Where it is possible to determine abundances using both sets, it has been found that chemical abundances estimated from CELs and ORLs yield systematically different values, with those from ORLs always being higher.
    This difference is quantified by means of the Abundance Discrepancy Factor (ADF) \citep{liu01}:
    \begin{equation}
        {\it ADF}({X}^{+i}) = \frac{n({X}^{+i})_{\mathrm{ORL}}}{n({X}^{+i})_{\mathrm{CEL}}},
    \end{equation}
    for ion $X^{+i}$.

    \ion{H}{ii} regions typically exhibit moderate ADF values, around 2 \citep[e.g.][]{garciarojas07,penaguerrero2012}; while in PNe moderate (2--10) and extreme values have been found, some close to 100 \citep[e.g.][]{mcnabb13}.

    A number of physical scenarios have been proposed to explain the origin of the abundance discrepancy: thermal inhomogeneities \citep{peimbert67,peimbert69}; coexistence of multiple gas phases with different chemical composition \citep{liu00}, notably the case of enrichment by ejecta from central binary stars in PNe \citep{corradi15, wesson18}; non-Maxwellian electron energy distributions \citep{nicholls12,nicholls13}; and varying ionizing radiation fields from short period binary stars \citep{bautista18}. Despite these important advances, none of the proposed scenarios has fully reproduced the observed behavior of emission lines and hence the origin of the abundance discrepancy remains an open problem in contemporary astrophysics \citep{garciarojas19}.

    Traditionally, CELs have been used to estimate the chemical composition of both \ion{H}{ii} regions and PNe, however the discovery of large ADFs has demanded a rethinking of this approach.
    
    For chemically homogeneous photoionized regions (e.g. almost all known \ion{H}{ii} regions), ORLs could provide a better representation of the chemical composition in the presence of small thermal inhomogeneities, since ratios of recombination lines are nearly temperature independent \citep{peimbert13}; meanwhile, CEL abundances are based on CEL to ORL line ratios which show a strong dependence on temperature.
    
    Nebular abundances play an important role in other fields of astrophysics, such as galactic formation and calibration of strong line methods. For this reason, it is important to find the average or representative chemical composition of a photoionized region, if this can be defined.

    In the case of PNe, small ADF values ($\le$5) can be explained by the presence of thermal inhomogeneities \citep{peimbert14}; while larger ADFs hint to the presence of chemical inhomogeneities.

    It is very likely that the ADF may have a different origin in every object. The real cause may be due to a contribution of most of the proposed physical scenarios. Nonetheless, it is clear that in order to provide insight into this problem, the detection of CELs and ORLs is desired. Moreover, it may be necessary to develop a detailed photoionization model of the object in question. 

    M 2-36 is a planetary nebula from the galactic bulge whose ADF was measured to be approximately 5 by \citet{liu01}. In that work, M 2-36 was studied with medium spectral resolution, using ESO's 1.52 m telescope; effectively resolving ORLs of heavy elements, including C$^{++}$, N$^{++}$, and O$^{++}$. Despite being groundbreaking at the time, recent works on PNe have highlighted the necessity of studying the gas phase with the highest spectral and spatial resolution possible in order to elaborate on the nature of the Abundance Discrepancy Problem. Hig spectral resolution (with a considerable signal-to-noise ratio) is desirable to detect and fully resolve ORLs, and to compute abundances and physical conditions from them \citep{peimbert14, mcnabb16}. Intermediate spectral resolution combined with  good spatial resolution and  detailed analysis (including the use of line-fitting algorithms) has proven to be effective in identifying and characterizing the volumes producing the emission from ORLs and comparing them with CEL emission, providing insight into the nature of the emitting material \citep{corradi15, garcia-rojas16, wesson18}. 

    In this work, we present a deep spectral analysis of M 2-36 based on observations from the Ultraviolet and Visual Echelle Spectrograph (UVES), installed at the ESO's 8.2 m Very Large Telescope (VLT). As we will show, the high resolution and spectral coverage ranging from the optical to the near infrared (3030 -- 10360 \AA) allows the detection of almost 450 emission lines, and a precise calculation of $T_{e}$ and $n_{e}$, which in turn yield chemical abundances with small error bars. The detection of multiple ORLs  from heavy elements with a high signal-to-noise ratio has made possible a recalculation of the ADF with small error bars, also allowing us to estimate the physical conditions in the nebula from this set of lines. The latter is important to test the hypothesis that the bulk of the emission of ORLs may arise from cold hydrogen-poor (metal-rich) clumps in the gas phase.

    In Section \ref{sec:observations} we describe the observations and data reduction. Line identification and extinction correction is presented in Section \ref{sec:lines}. In Section \ref{sec:phys_conditions} we present physical conditions. In Section \ref{sec:chem_abundances} we compute ionic and total abundances. In Section \ref{sec:inhomogeneities} we discuss the presence of inhomogeneities in the nebula, leading us to discuss photoionization models in Section \ref{sec:model}. Finally, in Section \ref{sec:conclusions} we present our summary and conclusions.

\section{Observations}
\label{sec:observations}
	
\begin{figure}
		\begin{center}
			\includegraphics[width=\columnwidth]{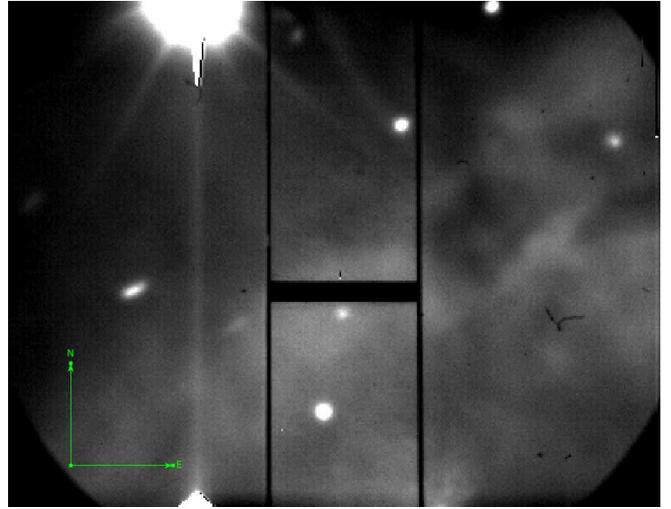}
		\end{center}
		\caption{Slit position across M 2-36. Its dimensions are 2 $\times$ 10
			arcsec$^{2}$.}
		\label{fig.slit}
	\end{figure}

    The observations were carried out during the night of march 30, 2003 (programme ID 70.C-0008(A)), using the Ultraviolet and Visual Echelle Spectrograph (UVES; \cite{dodorico00}) installed at the Very Large Telescope in Cerro Paranal, Chile. We observed the object simultaneously with the red and blue arms in two settings covering effectively the range from 3030 to 10360 \AA, except for small gaps resulting from the separation between the two CCDs used in the red arm. In all, unusable wavelength ranges were 5780--5830, 8530--8640, 10084--10090, and 10250 -- 10360 \AA. The slit dimensions were set to 2\arcsec $\times$ 10\arcsec, with a East-West orientation, effectively covering M 2-36 (Fig. \ref{fig.slit}) while avoiding the central star. The resolving power with this configuration was $\Delta \lambda/\lambda \sim$ 30000. The atmospheric dispersion corrector was used to keep the same observed region within the slit, regardless of the airmass value. Data were acquired according to the observation plan detailed in Table \ref{tab:journal_obs}.
	
	The spectra were reduced using IRAF\footnote{IRAF is distributed by NOAO, which is operated by AURA, Inc., under cooperative agreement with NSF.} following the standard procedure of
	bias subtraction, aperture extraction, flat fielding, wavelength calibration and
	flux calibration. For the latter, standard
	stars EG$\, 274$ and CD-329927 were observed. 
	
	\begin{table*}
		\label{tab:journal_obs}
		\caption{Journal of observations}
		\begin{tabular}{c c c c c} \hline
			Cross Disperser & Dichroic & $\lambda$ (\AA) & Exp. time (s) & Identification\\ 
			\hline
			1               & 1        & 3100 -- 3880 & 5 $\times$ 225 & UV \\
			2               & 2        & 3730 -- 4990 & 3 $\times$ 675 & BLUE\\
			3               & 1        &   4760 -- 6840 &  3 $\times$ 225  &        GREEN
			/ RED \\
			4               & 2         & 6600 -- 10360  &  3 $\times$ 675 & NEAR IR /
			FAR IR \\    \hline
		\end{tabular}
	\end{table*}
	
\section{Line fluxes, identification and extinction correction}
\label{sec:lines}		

	The reduced spectrum was analyzed using IRAF's \emph{splot} routine. The flux of the emission lines was determined by integrating between two points over the local continuum estimated by eye. When dealing with line blends, a gaussian profile was fitted using \emph{splot}, considering the laboratory wavelength as reference, allowing the actual wavelength to shift while keeping the separation constant.
	
	Some lines exhibit a double peak due to nebular expansion; in these cases,
	the average wavelength of the two components was used for identification, and
	the corresponding flux was calculated by integrating both peaks. 446 emission
	lines were identified, based on previous
	identifications by \citet{liu01} and the Atomic Line List v2.05b21
	\citep{vanhoof18}. Table \ref{tab:linelist} presents the results of our
	identifications: in Column 1 we present the laboratory
	wavelength $\lambda_{0}$ for air; Column 2 presents the ion emitting the line;
	Column 3 shows the multiplet originating the emission; Column 4 shows the flux
	corrected for reddening relative to
	H($\beta$), $I$(H$\beta$). The second to last column includes the fractional error
	(1$\sigma$) of the line intensities in percentage. Dubious identification is noted with a question mark.
	
\begin{figure}
	    \centering
	    \includegraphics[width=\columnwidth]{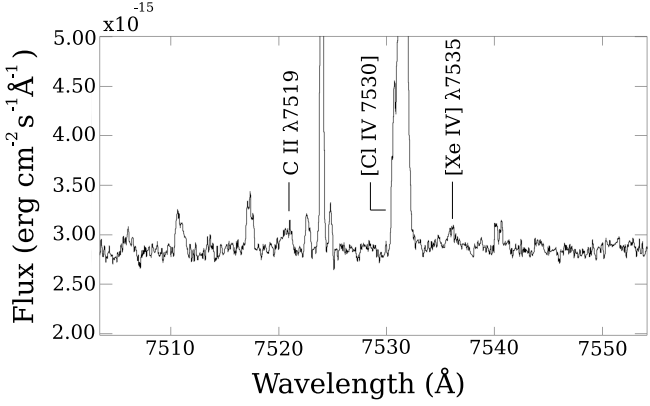}
	    \caption{[\ion{xe}{iv}] detected in our spectra, emitted by an ion created from the s-process.}
	    \label{fig:xe_iv}
	\end{figure}

	Notably we have resolved over forty recombination lines of \ion{O}{ii}, including the entirety of multiplet V1, which is central to study the ADF(O$^{++}$). We have also  identified three emission features corresponding to [\ion{Kr}{iv}], and [\ion{Xe}{iv}], (Figure \ref{fig:xe_iv}) which are produced by the s-process (slow neutron capture process); making these detections the first ones of neutron capture lines in PN M 2-36.
	
	The theoretical intensity ratios for the Balmer and Paschen emission lines were
	calculated using the program INTRAT by \cite{storey95} considering a constant
	electron temperature $T_{e}=8000$ K, and
	an electron density $n_{e}=5000$ cm$^{-3}$; there was no need to modify these
	values since hydrogen lines are nearly independent of temperature and density. 
	These theoretical values were compared to
	our measurements in order to estimate the extinction correction.

	We adopted the extinction law of \cite{cardelli89}. The logarithmic extinction correction for H$\beta$, $C$(H$\beta$), and the ratio of total to selective extinction, $R_{V}$, were fitted simultaneously and determined by reducing the quadratic
	discrepancies between the theoretical and measured H lines in units of the
	expected error, $\chi^{2}$. The fluxes were
	normalized with respect to the Balmer decrement, meaning that the value of
	$I$(H$\beta$) was allowed to deviate slightly from 100. The best fit parameters
	correspond to $C$(H$\beta)$=0.33 $\pm$ 0.01 and
	$R_{V}=2.75$. The error contribution from flux calibration was estimated to be
	1.5{\%}.

	The line fluxes where extinction corrected according to
	\begin{equation}
	I(\lambda) = 10^{C(\text{H}\beta)f(\lambda)}F(\lambda).
	\end{equation}
	Where $f(\lambda)$ is the extinction curve for $R_{V}=2.75$. 
	
\section{Physical conditions}
\label{sec:phys_conditions}	
	\subsection{Physical conditions from CELs}\label{subsec:physconditions}
    \label{subsec:phys_conditions_cel}
	
\begin{figure}
		\includegraphics[width=\columnwidth]{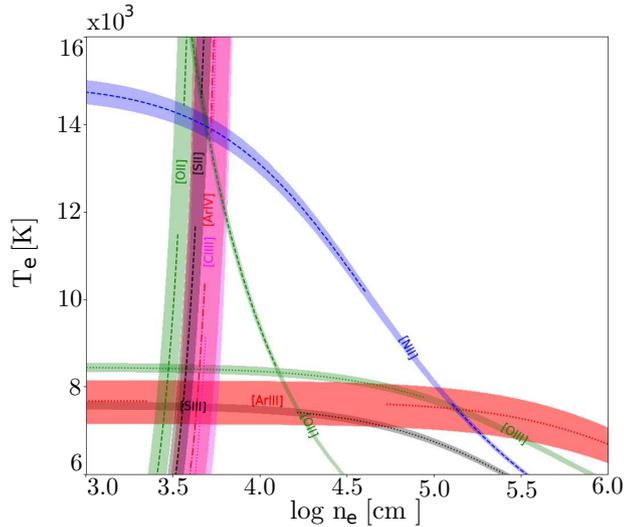}
		\caption{Electron temperature and density diagnostics from forbidden line
			ratios.}
		\label{fig:diags}
	\end{figure}

	Electron temperatures and densities obtained from various plasma diagnostics
	are presented in Table \ref{tab:temden}. We have made the distinction between
	high-ionization and low-ionization species,
	and assigned a respective electron temperature and density for each. Individual
	diagnostics were computed using PyNeb's \emph{getCrossTemden} routine \citep{luridiana15}, allowing a simultaneous determination of $T_{e}$ and $n_{e}$; the results are plotted in Figure \ref{fig:diags}. For [\ion{O}{iii}] we have considered the nebular to auroral line ratio $\lambda$4959/$\lambda$4363,
	excluding [\ion{O}{iii}] $\lambda$5007 since it was slightly saturated in our spectra. The adopted $T_{e}$ and $n_{e}$ for each ionization zone were obtained from the weighted average of the diagnostics considered. The atomic data set used is listed in Table \ref{tab:atomic}.

	Errors associated to $T_{e}$ and $n_{e}$ were propagated from intensity errors
	using Monte Carlo simulations. We generated 500 random values for each observed
	line intensity, considering a Gaussian
	distribution centered on said intensity. We verified that for larger random samples, the errors
	remained stable.

	\cite{escalante12} have pointed that \ion{O}{ii} emission lines, such as those
	from multiplet V1, may be affected by fluorescence in low excitation planetary
	nebulae such as IC 418. We have considered these effects to be negligible in M 2-36 since its ionization degree is much higher than that of IC 418. 
	
	[\ion{N}{ii}] $\lambda$5755 is known to be affected by recombination emission. Thus, the intensity was corrected using Equation 1 from \cite{liu00},  amounting to 3\%. [\ion{O}{ii}] $\lambda\lambda$7320+7330 is affected by recombination too; using Equation 2 from \cite{liu00} we find the recombination contribution to be 7\%. Both sets of lines were corrected before computing the physical conditions reported in Table \ref{tab:temden}.

	Our calculations are consistent with those reported by \cite{liu01} in their study of M 2-36, except for the case of [\ion{N}{ii}] $\lambda\lambda$6548+6583/$\lambda$5755. This difference cannot be attributed to the the recombination contribution to the lines of [\ion{N}{ii}]  since it is very small. It must be noted that our $T_{e}$([\ion{N}{ii}]) is consistent with our $T_{e}$([\ion{O}{ii}]) determination within 2$\sigma$, while the temperatures by Liu et al. are not consistent; on the other hand, our low-ionization temperatures are much higher than our high-ionization temperatures, while only the $T_{e}$([\ion{O}{ii}]) is high in that work. We must also note that our measured flux for [\ion{N}{ii}] $\lambda$5755 is considerably higher than the one reported by Liu et al. Another possible reason for the discrepancy is that it arises from the fact that we are observing different volumes within the nebula: since we observe an area of 2\arcsec $\times$ 10\arcsec and Liu et al. study the whole nebula we do not expect all quantities to be equal. In any case, the available data does not allow us to go deeper into the causes of the difference in $T_{e}$([\ion{N}{II}]).

	\begin{table*}
		\centering
		\caption{Temperatures and densities from forbidden line ratios}
		\label{tab:temden}
		\begin{tabular}{c c c c} 
			\hline
			Diagnostic                          									 & $T_{e}$ (K) & Diagnostic     
			& $n_{e}$ (cm$^{-3}$)\\
			\hline      
			\multicolumn{4}{c}{High Ionization}	\\ 
			{[\ion{O}{iii}]} $\lambda$4959/$\lambda$4363             		 		 &   $8380\pm100$           &   {[\ion{Cl}{iii}]} $\lambda$5517/$\lambda$5537		 & 		$6530\pm1800$					\\
			
			{[\ion{S}{iii}]} $\lambda\lambda$9069 + 9532/$\lambda$6312     		 &    $7600\pm150$          &   {[\ion{Ar}{iv}]} $\lambda$4711/$\lambda$4740 											 &   $4900\pm1100$ \\
			
			{[\ion{Ar}{iii}]} $\lambda\lambda$7136 + 7751/$\lambda$5192    		 &    $7650\pm500$  &   & \\
			{\bf Adopted}															 &	$8130\pm100$			&					& $5400\pm1000$ \\
			
			\multicolumn{4}{c}{Low ionization}				\\	
			
			{[\ion{N}{ii}]} $\lambda\lambda$6548 + 6583/$\lambda$5755       		 &    $14200\pm300$   &   {[\ion{O}{ii}]} $\lambda$3726/$\lambda$3729                           &  $3700\pm600$\\     
			
			{[\ion{O}{ii}]} $\lambda\lambda$3726 + 3729/$\lambda\lambda$7320 + 7330 &  $15100\pm400$ &   {[\ion{S}{ii}]} $\lambda$6716/$\lambda$6731 & $3400\pm500$\\
			{\bf Adopted}      						&      $14500\pm300$          &  					& $3500\pm600$\\
			\hline
		\end{tabular}
	\end{table*}
	
	\subsection{Physical conditions from ORLs}
	\label{subsec:phys_conditions_orl}
	Electron temperatures and densities can be obtained from ORLs of H$^{+}$, He$^{+}$ and O$^{++}$. 
	
\begin{figure}
		\includegraphics[width=\columnwidth]{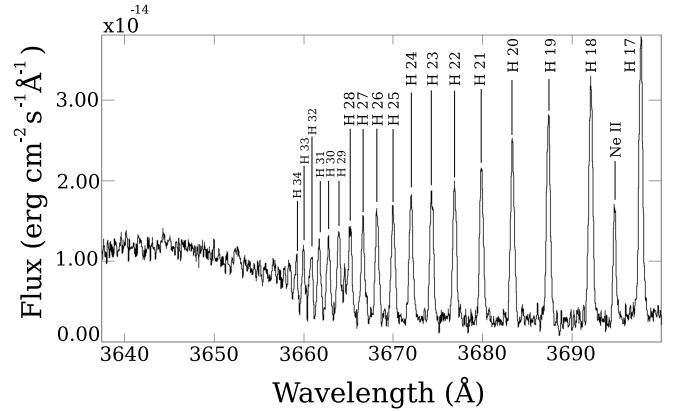}	
		\caption{Balmer decrement from our observations.}
		\label{fig.balmer}
	\end{figure}

	$T_{e}$(H$^{+}$) is usually obtained from the Balmer Jump discontinuity. In Figure \ref{fig.balmer} we show the spectrum near the discontinuity. By fitting gaussian profiles to the spectrum, using IRAF's
	\emph{splot} routine, we have resolved \ion{H}{i} recombination lines up to H36.
	The electron temperature can be computed from Equation 3 of 
	\cite{liu01},
	\begin{equation}
	T_{e} = 368 \times (1+0.259y^{+} + 3.409 y^{++})\left(
	\frac{\text{BJ}}{\text{H11}}\right)^{-3/2} \text{K},
	\end{equation}
	yielding a value of 6100 $\pm$ 900 K, which is in agreement with
	the previous determination by \cite{liu01}. We will refer to this value as $T_{e}$(BJ).
	
    $T_{e}$ can be obtained from the ratio of \ion{O}{ii} to [\ion{O}{iii}] lines, while the ratio of \ion{O}{ii} $\lambda$4649/V1 yields $n_{e}$ . Comparing the intensity of multiplet V1 of \ion{O}{ii} to [\ion{O}{iii}] $\lambda$4959 and using equation (1) from \cite{peimbert14}, we find that $T_{e} =$ 5400 $\pm$ 400 K. We also estimated $n_{e}$ from Figure 3 of \cite{peimbert13},  obtaining $n_{e}=$ 1100 $\pm^{1000}_{600}$ cm$^{-3}$. We will refer to these quantities as $T_{e} $(V1/4959) and $n_{e}$(V1).
    
    \ion{He}{i} lines can provide an estimation of $T_{e}$ and $n_{e}$. From the intensity ratio of \ion{He}{i} $\lambda$7281/$\lambda$6678, and Figure 2 of \cite{zhang05}, we estimate  $T_{e}$(\ion{He}{i})$=$ 5400 K. Similarly, we have estimated $T_{e}$ using HELIO14 \citep{peimbert12} and found $T_{e}$(\ion{He}{i})$=$7500 K (note that for this estimation, we have to set $t^{2}=0.00$ in the code). Note that in the work of Zhang, the authors find a value of 2790$\pm$1000 K. They also find $T_{e}$(\ion{He}{i}) to be systematically lower than $T_{e}({\rm BJ})$ in all but one object out of a sample of fifty PNe. Our result thus contradicts the one reported by \cite{zhang05}; this will be explored further in Section \ref{sec:model}.
    
    Physical conditions from ORLs are summarized in Table \ref{tab:ORL_phys}.
    
    It is remarkable that temperatures derived from heavy element ORLs are systematically lower than those derived from CELs. This is consistent with the results obtained by \cite{peimbert14} in PNe, however the difference between $n_{e}$(V1) appears to be significantly larger than the average found in their sample. $n_{e}$(V1) is irreconcilable with $n_{e}$[\ion{Cl}{iii}] and any other CEL diagnostic; this will be addressed in section \ref{sec:model}. 
    
	\begin{table}
	    \centering
	    \caption{Physical conditions from ORLs}
	    \begin{tabular}{c c c} \hline
	       Method & $T_{e}$ (K) & $n_{e}$ (cm$^{-3}$) \\ \hline
	       \ion{H}{i} Balmer Jump, $T_{e}$(BJ)      &  6100 $\pm$ 900   &                       \\
	       \ion{He}{i} $\lambda$7281/$\lambda$6678  &  5400             &                       \\
	       \ion{He}{i} (HELIO 14)                   &  7500             & 6380  $\pm$ 1500 \\
	       \ion{O}{ii} (V1)                         &                   & 1100 $\pm^{1000}_{600}$ \\
	       \ion{O}{ii}(V1/$\lambda$4959)            & 5400 $\pm$ 400    &               \\ \hline
	    \end{tabular}
	    \label{tab:ORL_phys}
	\end{table}
	
\section{Chemical Abundances}
\label{sec:chem_abundances}

    \subsection{Ionic abundances from collisionally excited lines}
    \label{subsec:cel_abund}
	
	We have computed the ionic abundances from multiple CELs of heavy
	elements. All computations were performed with PyNeb considering the atomic data from
	Table \ref{tab:atomic}. Errors were propagated
	from those associated to line intensities using Monte Carlo simulations. As
	explained in section \ref{subsec:physconditions}, we have used a two-zone
	scheme, distinguishing between species originating mainly in the high ionization zones (O$^{++}$, Ne$^{++}$, S$^{++}$, Cl$^{++}$, Cl$^{3+}$, Ar$^{++}$, Ar$^{3+}$, Kr$^{3+}$, and Xe$^{3+}$) and those that originate in the low
	ionization zones (N$^{0}$, N$^{+}$, O$^{0}$, O$^{+}$, S$^{+}$ and Cl$^{+}$). Our results are presented in Table \ref{tab:abund_cel}.
	
	\begin{table*}
	\caption{Atomic data set used for collisionally excited lines.}
	    \label{tab:atomic}
	    \centering
	    \begin{tabular}{c c c} \hline
	       Ion  & Transition probabilities & Collisional strengths \\ \hline 
	       N$^{0}$      &   \cite{wiese96}          &  \cite{pequignot76}  \\
	        N$^{+}$    &           \cite{fischer04}           &          \cite{tayal11}    \\ 
	       O$^{0}$      &   \cite{wiese96}          &  \cite{pequignot76}  \\
	        O$^{+}$     &       \cite{fischer04}        &   \cite{kisielius09} \\
	        O$^{++}$    &   \cite{fischer04}        &   \cite{storey14} \\
	                    &   \cite{storey00}        &            \\
	        Ne$^{++}$   &   \cite{galavis97}    &   \cite{mclaughlin00} \\
	        S$^{+}$     &   \cite{podobedova09}    &  \cite{tayal10} \\
	        S$^{++}$    & \cite{podobedova09}   & \cite{tayal99} \\
	        Cl$^{+}$    & \cite{mendoza83} & \cite{tayal04} \\
	        Cl$^{++}$   & \cite{mendoza83} &  \cite{butler89} \\
	        Cl$^{3+}$   & \cite{kaufman86} &  \cite{galavis95} \\
	                    & \cite{mendoza82a} &                   \\
	                    & \cite{ellis84}    &                   \\
            Ar$^{++}$   &  \cite{mendoza83} & \cite{galavis95} \\
	                    &   \cite{kaufman86} &                  \\
	        Ar$^{3+}$   & \cite{mendoza82b} & \cite{ramsbottom97} \\
	        Kr$^{3+}$   & \cite{biemont86a} & \cite{schoning97} \\ 
	        Xe$^{3+}$   &   \cite{biemont95} & \cite{schoening98} \\ \hline                                        
	    \end{tabular}
	\end{table*}
	
	\begin{table}
		\centering
		\caption{Ionic abundances from collisionally excited lines}
		\label{tab:abund_cel}
		\begin{tabular}{l c c c}\hline
			Ion         &   $t^{2} =$ 0.00  &           &       \\   \hline
			N$^{0}$		&	5.59 $\pm$ 0.05 &			&			\\
			N$^{+}$     &   7.10 $\pm$ 0.04 &           &       \\
			O$^{0}$		&	6.60 $\pm$ 0.03	&			& 			\\
			O$^{+}$     &   6.96 $\pm$ 0.04 &           &       \\
			O$^{++}$    &   8.79 $\pm$ 0.02                  &           &       \\
			Ne$^{++}$   &    8.30  $\pm$ 0.03            &           &       \\
			S$^{+}$     &   5.64 $\pm$ 0.04                  &           &       \\
			S$^{++}$    &   6.92 $\pm$ 0.03                   &           &       \\
			Cl$^{+}$    &   3.84 $\pm$ 0.08                   &           &       \\
			Cl$^{++}$   &   5.28 $\pm$ 0.05                  &           &        \\
			Cl$^{3+}$   &   4.72 $\pm$ 0.04    &           &       \\
			Ar$^{++}$   &   6.48 $\pm$ 0.04                   &           &        \\
			Ar$^{3+}$   &   5.97 $\pm$ 0.04                   &           &        \\
			Kr$^{3+}$   &   3.30 $\pm$ 0.11 & 				&			\\
			Xe$^{3+}$   &   2.70 $\pm$ 0.15 & 				& 		\\ \hline
		\end{tabular}
	\end{table}
	
    \subsection{Ionic abundances from recombination lines}
    \label{subsec:abund_orl}
    \begin{table}
        \caption{Atomic data used for recombination lines.}
        \label{tab:atomic_rec}
        \centering
        \begin{tabular}{c c} \hline
          Ion   & Recombination coefficients  \\ \hline
            H$^{+}$ & \cite{storey95} \\
            He$^{+}$ & \cite{porter12, porter13} \\
            He$^{++}$ &  \cite{storey95} \\
            C$^{++}$ & \cite{davey2000} \\
            N$^{++}$ & \cite{fang11} \\ 
            O$^{++}$ & \cite{storey17} \\
            Ne$^{++}$ & \cite{kisielius98} \\ \hline
        \end{tabular}
    \end{table}

	We have detected 58 \ion{He}{i} and 8 \ion{He}{ii} lines in our spectrum. Most of the
	helium found in the observed volume is singly ionized.
	
	The He$^{+}$/H$^{+}$ ratio was computed from 9 lines using the package HELIO14, an update of the software  described in \citet{peimbert12}. This package uses a maximum likelihood method
	to fit simultaneously He$^{+}$/H$^{+}$;
	electron density; the optical deph of \ion{He}{i} $\lambda$3889, $\tau_{3889}$;
    the normalized mean square thermal inhomogeneity, $t^{2}$; and $T_{0}$. One of the main
	advantages of this software is the precise
	calculation it produces, with an accuracy of up to three significant figures. Results are presented in Table \ref{tab:he_ab}.
	
	 For He$^{++}$/H$^{+}$,
	the only two lines not contaminated by emission
	from other ions that present errors lower than 15\% are \ion{He}{ii} $\lambda$4686 and
	\ion{He}{ii} $\lambda$3203. We have adopted the value derived from \ion{He}{ii} $\lambda$4686 since it is
	the strongest one.

	\begin{table}
	    \caption{Ne$^{++}$/H$^{+}$ ionic abundances from optical recombination lines}
	    \label{tab:Ne_rl}
	    \centering
	    \begin{tabular}{c c} \hline
	        $\lambda_{0}$      &  12+log(Ne$^{++}$/H$^{+}$)\\ \hline
	       3694.21      &  9.02 $\pm$ 0.06 \\
	        3709.62     &   8.98 $\pm$ 0.08 \\
            3766.26	    &   8.60: \\
	        3777.14     &   8.82: \\ 
	        {\bf Adopted} & 9.00 $\pm$ 0.08 \\ \hline 
	    \end{tabular}
	\end{table}
	
	\begin{table}
	    \caption{N$^{++}$/H$^{+}$ abundances from recombination lines}
	    \label{tab:N_rl}
	    \centering
	    \begin{tabular}{c c c} \hline
	    $\lambda_{0}$   & Mult.   &  12+log(N$^{++}$/H$^{+}$)\\ \hline
	       4035.08  & V39a & 9.26 $\pm$ 0.08\\
	       4041.31  & V39b & 9.21 $\pm$ 0.06 \\
	       4043.53  & V39a & 9.11 $\pm$ 0.10 \\
	       4607.16  & V5 & 8.84 $\pm$ 0.14 \\ 
	       4621.39  & V5 & 9.06 $\pm$ 0.11 \\
	       4630.54  & V5 & 9.16 $\pm$ 0.05 \\ 
	       4643.09  & V5 & 8.78 $\pm$ 0.13 \\
            4788.13 & V20 & 8.64 $\pm$ 0.17 \\
            5495.67 &  3P--3P$_{0}$& 8.70 $\pm$ 0.16 \\
            5666.63 & V3 & 9.14 $\pm$ 0.05 \\ 
            5676.02 &  V3& 9.09 $\pm$ 0.08 \\
            5679.56 & V3 & 9.20 $\pm$ 0.03 \\
            5686.21 & V3 & 9.14 $\pm$ 0.10  \\
            5710.77 &  V3 & 9.21 $\pm$ 0.08 \\
            5927.81 & 3P--3D$_{0}$ & 8.67 $\pm$ 0.11 \\
	        {\bf Adopted} & & 9.16 $\pm$ 0.09 \\ \hline
	    \end{tabular}
	\end{table}

	\begin{table}
	    \caption{He$^{+}$/H$^{+}$ and He$^{++}$/H$^{+}$ from ORL}
	    \label{tab:he_ab}
	    \centering
	    \begin{tabular}{c c} \hline
	        $\lambda_{0}$ & 12 + log(He$^{+}$/H$^{+}$)  \\ \hline
	        3819 & 11.091 \\
            3889 & 11.101 \\
            4388 & 11.089 \\ 
            4471 & 11.125 \\
            4713 & 11.114 \\ 
	        4922 & 11.068 \\
	        5876 & 11.083 \\
	        6678 & 11.144 \\
	        7065 & 11.102 \\
	         \hline
	        {\bf Adopted} & 11.101 $\pm$ 0.004 \\
	        \hline\hline
	        $\lambda_{0}$ & 12 + log(He$^{++}$/H$^{+}$) \\ \hline
 	          {\bf4686 (adopted)} & 9.52 $\pm$ 0.02 \\ \hline
	    \end{tabular}
	\end{table}

    Besides \ion{H}{I}, \ion{He}{I}, and \ion{He}{II}, we have detected several ORLs produced by C$^{++}$, N$^{++}$, O$^{++}$ and Ne$^{++}$. We have computed ionic abundances for all of them considering the high ionization $T_{e}$ and $n_{e}$, and the atomic data from Table \ref{tab:atomic_rec}; they are presented in Tables \ref{tab:Ne_rl}, \ref{tab:N_rl}, \ref{tab:c_abund} and \ref{tab:oxygen_orl_abund}.  For Ne$^{++}$  we adopted the weighted average of $\lambda$3694.21 and $\lambda$3709.62; for N$^{++}$ we  adopted the weighted average from multiplet V5 lines.

\begin{figure}
		\includegraphics[width=\columnwidth]{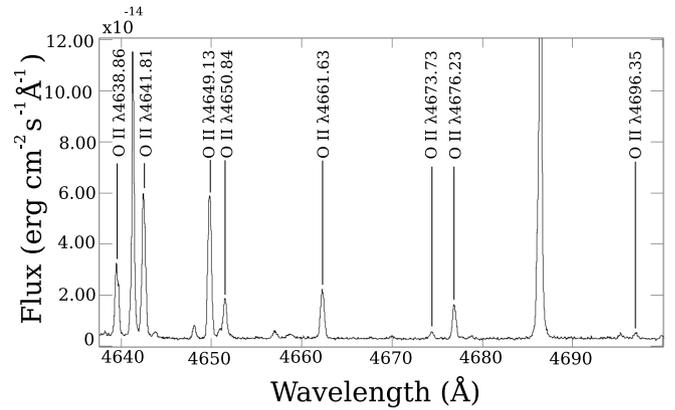}
		\caption{Spectrum around multiplet V1 of O II}
		\label{fig:multV1}
	\end{figure}

    Special attention has been dedicated to recombination lines of \ion{O}{ii}, of which we have detected all lines that make up multiplet V1, namely \ion{O}{ii} $\lambda$4638.86, $\lambda$4641.81, $\lambda$4649.13, $\lambda$4650.84, $\lambda$4661.63, $\lambda$4673.73, $\lambda$4676.23, and $\lambda$4696.35. All lines were detected with a considerable signal-to-noise ratio allowing us to compute the ionic abundance with small errors (Table \ref{tab:oxygen_orl_abund}). We have adopted the abundance obtained from the weighted average of multiplet V1. In Figure \ref{fig:multV1} we show the quality of the spectrum around this multiplet. 

    \ion{O}{ii} lines from other multiplets are also found in our spectrum. We have selected those with reasonable errors (no greater than 30\%) in order to compute O$^{++}$/H$^{+}$, also we have been careful not to include lines that may be contaminated by those from other ions. 
	
	Table \ref{tab:oxygen_orl_abund} highlights that 3d-4f transitions yield
	systematically larger values for O$^{++}$/H$^{+}$ than those derived from
	multiplet V1. \cite{peimbert13} have argued that \ion{O}{ii} $\lambda$4089.29 may be
	contaminated by \ion{Si}{iv} $\lambda$4088.86, meaning the value derived for
	O$^{++}$ from this line could be overestimated.
	In our spectrum, \ion{Si}{iv} $\lambda$4088.86 should be present at $\lambda$4089.30; 
	we have verified that there is a feature present on the blue side of \ion{O}{ii} $\lambda$4089.29. While the wavelength is consistent with \ion{Si}{iv} $\lambda$4088.86, recent work by \cite{mendezdelgado21} shows that the feature registered by \cite{peimbert13} is also consistent with an artifact arising from internal reflections in dichroic \#2 in the blue arm of UVES. On the other hand, we identify a weak feature at $\lambda$4116.74 which is produced by the same multiplet of \ion{Si}{iv}, strengthening the possibility of the presence of \ion{Si}{iv} $\lambda$4088.86; neither of the features can be measured well enough to determine their ratio with any confidence nor is it possible to determine the abundance of Si$^{+4}$ from \ion{Si}{iv} $\lambda$4116.74. Therefore we cannot establish firmly whether the feature at $\lambda$4089.30 is produced by \ion{Si}{iv} or an internal reflection.
	
	For the case of C$^{++}$,  \cite{grandi76} recommends using the lines that populate \ion{C}{ii} $\lambda$4267.15, therefore we have considered the weighted average of the abundances listed in Table \ref{tab:c_abund}. We have used the recombination coefficients by \cite{davey2000}, considering case B recombination for the temperature range $T_{e}=$ 2500--30000 K, and electron density $n_{e} = 10^{4}$ cm$^{-3}$. Our values agree with the results obtained by \cite{liu01}.

	\begin{table}
		\begin{center}
			\caption{C$^{++}$/H$^{+}$ from optical recombination lines} 
			\label{tab:c_abund}
			\begin{tabular}{c c c} \hline
				$\lambda_{0}$	& Mult. & 12 $+$ log(C$^{++}$/H$^{+}$) \\
				\hline
				4267.15 & 6     &  9.34 $\pm$ 0.06 \\
				5342.38 & 17.06 & 9.32 $\pm$ 0.08 \\
				6151.43 & 16.04 & 9.22 $\pm$ 0.06 \\
				6461.95 & 17.04 & 9.36 $\pm$ 0.04 \\
				9903.46 & 17.02 & 9.45 $\pm$ 0.03 \\
				\multicolumn{2}{c}{{\bf Adopted}} & 9.36 $\pm$ 0.03 \\ \hline
			\end{tabular}
		\end{center}		
	\end{table}

	\begin{table}
		\begin{center}
			\caption{O$^{++}$/H$^{+}$ ionic abundances from optical recombination lines}
			\label{tab:oxygen_orl_abund}
			\begin{tabular}{c c c}
				Mult. & $\lambda_{0}$ & 12 $+$ log(O$^{++}$/H$^{+}$) \\
				\hline
				V1	& 4638.86	& 9.75 $\pm$ 0.04	\\
				V1	& 4641.81	&	9.72 $\pm$ 0.02 \\
				V1	& 4649.13	&	9.56 $\pm$ 0.03	\\
				V1	& 4650.84	&	9.53 $\pm$ 0.04	\\
				V1	& 4661.63	& 9.57 $\pm$ 0.04	\\
				V1	& 4673.73	& 9.48 $\pm$ 0.11	\\
				V1	& 4676.23	& 9.55 $\pm$ 0.05	\\		
				V1	& 4696.35	& 9.72 $\pm$ 0.12 	\\
				{\bf V1 sum (adopted)} &	& 9.62 $\pm$ 0.03	\\
				\hline
				V2 & 4317.14 & 9.48 $\pm$  0.08 \\
				V2 & 4319.63 & 9.52 $\pm$  0.08 \\
				V2 & 4325.76 & 9.49 $\pm$  0.15 \\
				V2 & 4345.56 & 9.43 $\pm$  0.07 \\
				V2 & 4349.43 & 9.55 $\pm$  0.14 \\
				V2 & 4366.89 & 9.54 $\pm$  0.10 \\
				{\bf V2 sum} & &9.50	$\pm$ 05\\	
				\hline	
				V5 & 4414.90 & 9.47 $\pm$ 0.07 \\
				V5 & 4416.97 & 9.50 $\pm$ 0.08 \\
				{\bf V5 sum} & &9.48 $\pm$ 0.06\\	
				\hline
				V10 & 4069.88 & 9.74 $\pm$ 0.03 \\
				V10 & 4072.16 & 9.63 $\pm$ 0.03 \\
				V10 & 4075.86 & 9.76 $\pm$ 0.03 \\
				V10 & 4078.84 & 9.48 $\pm$ 0.10 \\
				V10 & 4085.11 & 9.52 $\pm$ 0.08 \\
				V10 & 4092.93 & 9.59 $\pm$ 0.09 \\
				{\bf V10 sum} &	&9.70 $\pm$ 0.05	\\
				\hline
				V19 & 4121.46 & 9.41 $\pm$ 0.10 \\
				V19 & 4132.80 & 9.40 $\pm$ 0.08 \\
				V19 & 4153.30 & 9.52 $\pm$ 0.06 \\ 
				V19 & 4156.53 & 10.03 $\pm$ 0.08 \\
				V19 & 4169.22 & 9.71 $\pm$ 0.08 \\
				{\bf V19 sum} &	&9.56 $\pm$0.05\\
				\hline
				V20 & 4110.79 & 9.67 $\pm$ 0.08 \\
				V20 & 4119.22 & 9.58 $\pm$ 0.05 \\
				{\bf V20 sum}&	& 9.60 $\pm$ 0.04		\\
				\hline
				V25 & 4699.22 & 9.64 $\pm$ 0.16 \\
				V25 & 4705.35 & 9.67 $\pm$ 0.14 \\
				{\bf V25 sum} & & 9.66 $\pm$ 0.11 \\
				\hline
				\multicolumn{3}{c}{{\bf 3d-4f transitions}}\\
				V48a & 4071.23 & 9.78 $\pm$ 0.09 \\
				V48a & 4089.29 & 9.74 $\pm$ 0.04 \\
				V48b & 4083.90 & 9.56 $\pm$ 0.08 \\
				V50a & 4062.94 & 10.08 $\pm$ 0.08 \\ 
				V53b & 4294.78,.92 & 9.55 $\pm$ 0.08 \\
				V86a & 4491.23 & 9.68 $\pm$ 0.09 \\
				V92a & 4609.44 & 9.74 $\pm$ 0.06 \\
				V92b & 4602.13 & 9.83 $\pm$ 0.07 \\
				V92c & 4610.20 & 10.03 $\pm$ 0.07 \\ 
				\hline
			\end{tabular}
		\end{center}
	\end{table}

	\subsection{Total abundances}
	\label{subsec:total_abund}
	In order to compute total elemental abundances, ionic abundances have to be corrected for the presence of unobserved ions in the gas. This is done by means of an Ionization Correction Factor (ICF). Total abundances for M 2-36 are reported in Table \ref{tab:total_abund}.
	
	For O, Ne, S, and Ar we used the ICFs developed by \cite{DIMS14} through photoionization models of PNe.
	
    Since we detect several \ion{He}{ii} lines and $\omega=\mathrm{O}^{++}/(\mathrm{O}^{+}+\mathrm{O}^{++})=0.98$, we have to account for the presence of O$^{3+}$ when computing O/H.  
	
	\cite{DIMS14} recommend not to use their ICF for N when $\omega \ge$ 0.95 and in turn we refer to the one proposed by \cite{kingsburgh94}, which assumes that N/O $=$ N$^{+}$/O$^{+}$. 
	
	We detect Cl$^{+}$, Cl$^{++}$ and Cl$^{3+}$ in our spectrum of M~2-36, therefore we can obtain Cl/H by adding their ionic abundances. We have also computed Cl/H using the ICF provided by \cite{liu00}; yielding a value smaller than the direct sum by 0.28 dex. We do not use \cite{DIMS14} ICF in this case since it is only valid when $0.02 \le \omega \le 0.95$. We have adopted the value obtained from the direct sum.
	
    Regarding neutron capture elements, only [\ion{Kr}{IV}] $\lambda$5867.74 and [\ion{Xe}{IV}] $\lambda$7535.40 are fully resolved, therefore we have considered them using equation (3) of \cite{sterling15}, from which we compute both Kr/H and Xe/H, given the similar ionization potential of Kr$^{3+}$ and Xe$^{3+}$.

    Abundances derived from observations are consistent with a Peimbert Type I PN, according to the criterium first stated by \cite{peimbert78} and later refined by \cite{peimbert83} that requires He/H $\ge$ 0.125 and N/O $\ge$ 0.5. \cite{kingsburgh94} proposed that Type I PNe can be defined from the ratio of N/O alone; based on recent photoionized region and solar abundances, \cite{henry04} have established a value of N/O $\ge$ 0.65 for Type I PNe. We obtain He/H $=$ 0.129 and N/O $=$ 1.44, implying that M 2-36 is a strong Type I PN.
    
	Our ORL abundances for He and C agree with those of \cite{liu01} and \cite{ratag97} within 1$\sigma$ (see Table \ref{tab:abund_compared}). Nonetheless, there is a significant difference between CEL abundances of O and N; our Ne/H determination agrees with that of Ratag et al., but not with Liu et al. This is to be expected because, as we will see in sections \ref{sec:inhomogeneities} and \ref{sec:model}, this object is composed from several notably different components; under such conditions it would be very unlikely to find uniform chemical abundance determinations between observations that do not include the emission from the entire object.

	\begin{table}
	    \centering
	    	    \caption{Total abundances.}
	    \begin{tabular}{c c c c} \hline
	       &  ICF                 &     CELs$^a$    &    ORL      \\ \hline
        	He & He$^{+}$ + He$^{++}$  &            & 11.112 $\pm$ 0.008    \\ 
        	C & Assuming C/O = C$^{++}$/O$^{++}$    &       & 9.36 $\pm$ 0.10 \\
        	N    &   \cite{kingsburgh94}   &     8.95 $\pm$ 0.06  & 9.17 $\pm$ 0.11 $^b$\\ 
        	O  &  \cite{DIMS14}       &     8.79 $\pm$ 0.03 & 9.64 $\pm$ 0.04 \\ 
	Ne      &\cite{DIMS14}       &     8.37 $\pm$ 0.06 & 9.02 $\pm$ 0.07\\
            & \cite{peimbert69}   &     8.36 $\pm$ 0.05  &\\
	S       &\cite{DIMS14}       &    7.35 $\pm$ $^{0.2}_{0.12}$ &\\
            &\cite{stasinska78}  &    6.94 $\pm$ 0.05 & \\
	Cl      &Cl$^{+}$ + Cl$^{++}$ + Cl$^{3+}$ & 5.41 $\pm$ 0.08 &\\
	        &\cite{liu00}        &   5.68 $\pm$ 0.10   &\\
	Ar      &\cite{DIMS14}       &    6.77 $\pm$ 0.7 &   \\
	Kr      & \cite{sterling15} &    3.44 $\pm$ 0.15 \\
	Xe      & \cite{sterling15} &   2.83: \\ \hline
	\multicolumn{4}{l}{$^a$ Abundances determined with $t^2=0.00$.} \\
	\multicolumn{4}{l}{$^b$ Assuming N/O = N$^{++}$/O$^{++}$ (ORLs).} \\
	\multicolumn{4}{l}{Colons indicate very uncertain abundances.}
		    \end{tabular}
	    \label{tab:total_abund}
	    
	\end{table}
	
	\begin{table*}
		\caption{Total abundances compared}
	    \centering
	    \begin{tabular}{c c c c c c c} \hline
	        &  \multicolumn{2}{c}{This work} & \multicolumn{2}{c}{Liu et al. (2001)} & Ratag et al. (1997) \\ \hline 
	        & CEL                       &     ORL &   CEL &   ORL     &   CEL       &       \\
	    He  &                           &  11.112 $\pm$ 0.008   &       &   11.13   &  11.09    \\
	     C  &                           &  9.36 $\pm$ 0.10      &  8.73 &  9.41     &   \\
	    N   & 8.95 $\pm$ 0.06           &   9.17 $\pm$ 0.11    &  8.42 &  9.17     & 8.62 \\
	    O   &  8.79 $\pm$ 0.03          & 9.64 $\pm$ 0.04       &  8.85 &  9.64     & 8.89\\
	    Ne  &  8.36 $\pm$ 0.05          & 9.02 $\pm$ 0.07       &  8.57 &  9.16     & 8.27\\
	    S   &  7.35 $\pm^{0.2}_{0.12}$  &                       &  7.47 &           & 7.17 \\   
	    Cl  & 5.41 $\pm$ 0.08            &                      &  5.42 &           & 5.98\\
	    Ar  &  6.77 $\pm$ 0.7            &                      &  6.66 &           & 6.61\\
	    Kr  &  3.44 $\pm$ 0.15           &                      &                   & \\
        Xe &    2.83:               &                           &                   &\\ \hline
	    \end{tabular}
	    \label{tab:abund_compared}
	\end{table*}
	
		\subsection{Abundance Discrepancy Factors} 
		\label{subsec:adf}
	
	Our ADF values are presented in Table \ref{tab:adf}.  For ADF(O$^{++}$), our value of 6.76 $\pm$ 0.50 is somewhat higher than the global ADF obtained from \cite{liu01}, taking \ion{O}{ii} V1 and optical [\ion{O}{iii}] abundances as reference, which is $\sim$5.50. Nonetheless, the ADF value can change depending on the methodology used. The small differences may be due to the fact that the two works use slightly different methodologies, and different zones of the object have been analyzed. Also we must note that previous works on the ADF in PN M 2-36 do not include error bars.
	
	Although it has been found that some PNe with ADF$\le$8 or similar can be chemically homogeneous, this is not a strict discriminator. Some of the parameters exhibited by M 2-36 suggest that it is not chemically homogeneous, notably, the irreconcilable values obtained for $t^{2}$ (further discussed in Section \ref{sec:inhomogeneities}), as well as the large difference in $T_{e}$ and $n_{e}$ between high and low ionization stages.
	
		\begin{table}	
	\caption{Abundance Discrepancy Factors. C$^{++}$ and N$^{++}$ ADFs were computed from our ORL abundances and the CEL abundances by \citet{liu01}}
	\label{tab:adf}
	\centering
	    \begin{tabular}{c c} \hline
	        Ion & ADF \\ \hline
	        C$^{++}$/H$^{+}$& 5.09 \\
	        N$^{++}$/H$^{+}$ & 7.30 \\
	        O$^{++}$/H$^{+}$ & 6.76 $\pm$ 0.50 \\
	        Ne$^{++}$/H$^{+}$ & 5.01 $\pm$ 0.52 \\ \hline
	    \end{tabular}
	\end{table}

\section{Inhomogeneities}
\label{sec:inhomogeneities}
	
    The existence of the ADF means that homogeneous single-phase models do not reproduce PNe adequately, implying the existence of inhomogeneities in either chemical composition, density, and/or temperature \citep[see][and references therein]{peimbert17}. Regarding the density inhomogeneities, while the filling factor is well known, \citep[see][and subsequent editions]{osterbrock74} it will not result in an ADF; in order to produce large ADFs, the inhomogeneities needed include clumps with $n_{e} \ge 100\,000$ cm$^{-3}$ as well as a surrounding volume with densities in the few thousands; in the remaining of this paper, this is what we will be referring to when we make reference to density inhomogeneites. Two important open questions remain: what is the relevance of each of the possible sources of inhomogeneity? and, what is the physical size of the relevant inhomogeneities?
	
	It must be noted that inhomogeneities in either density or chemical composition will also manifest themselves as thermal inhomogeneities. Inhomogeneities in density, when large enough, are able to distort our measurements by distorting nebular lines, which in turn affect the cooling and produce thermal inhomogeneities;  also the expected chemical inhomogeneities will affect the number of coolants in the gas, producing thermal inhomogeneities (and probably inducing density inhomogeneities). 
	
	An empirical way to tackle the thermal inhomogeneities is the $t^{2}$ formalism developed by \cite{peimbert67}. But the only way to study chemical and density inhomogeneities is to include different phases in photoionization models (a modified version of this strategy can also be used to study thermal inhomogeneities).
	
	\subsection{Thermal inhomogeneities}
	\label{subsec:thermal_inhomogeneities}

	The first evidence of a temperature structure in photoionized regions was reported by \cite{peimbert67} who found a considerable difference between $T_{e}$ derived from the Balmer Jump, and from forbidden line ratio diagnostics in the Orion Nebula. These differences are also found in PNe \citep[see, for example][]{zhang04}. This led to the development of the $t^{2}$ formalism to account for temperature variations, further developed in \cite{peimbert69}.
	
	To a second order approximation, the temperature structure of a photoionized region along the line of sight, can be characterized by the average temperature, $T_{0}$, and the normalized mean square
	temperature fluctuation, $t^{2}$, given by
	\begin{equation}
	T_{0}(\text{X}^{+i}) = \frac{\int T_{e}n_{e}n(\text{X}^{+i})dV}{\int
		n_{e}n(\text{X}^{+i})dV},
	\end{equation}
	and 
	\begin{equation}
	t^{2}(\text{X}^{+i}) = \frac{\int
		(T_{e}-T_{0}(\text{X}^{+i}))^{2}dV}{T_{0}(\text{X}^{+i})^{2} \int n_{e}
		n(\text{X}^{+i})dV},
	\end{equation}
	where $n_{e}$ is the electron density and $n(\text{X}^{+i})$ is the density of ion X$^{+i}$. Note that while $t^{2}$ represents the mean square deviations, there is no definition for the rms deviations, and thus $\sqrt(t^{2}) \ne t$.
	
	To determine $t^{2}$, two independent determinations of $T_{e}$ are required; preferably comparing a temperature that uses lines that favor the hottest parts of a nebula with one that uses lines that favor the cooler parts of the nebula; temperatures derived using optical CELs will better represent the hotter parts, while those derived using ORLs will represent the cooler parts. In this work, besides the temperature determination from forbidden line ratios, we have used the Balmer Jump; $T_{e}$ derived from helium ORLs, $T_e(\text{\ion{He}{i}}$); and a hybrid O temperature equivalent to the ADF(O$^{++}$). Our results for $t^{2}$ are summarized in Table \ref{tab:t2}.

	\begin{table}
		\caption{$t^{2}$ derived from different methods}
		\label{tab:t2}
		\begin{tabular}{c c c c} \hline
			Method 	&	$t^{2}$(\ion{O}{II})	&	$t^{2}$(BJ)	& $t^{2}$(\text{He}$^{+}$) \\
			Value	&	0.088 $\pm$ 0.003	&	0.048 $\pm$ 0.010	&	0.017 $\pm$ 0.003 \\ \hline
		\end{tabular}
	\end{table}
	
	It must be noted that the three $t^{2}$ values obtained in Table \ref{tab:t2} are irreconcilable. In other works, the $t^{2}$ values obtained from \ion{He}{i} lines have been found to agree with those obtained from the Balmer Jump, $t^{2}$ (BJ), and from the ADF(O$^{++}$), $t^{2}$(\ion{O}{ii}).  \cite{peimbert14}, for example find agreement in $t^{2}$(O$^{++}$) and $t^{2}$(He$^{+}$) in part of a sample of low ADF PNe; however this is not the case for M 2-36. When these three values agree we are led to consider a single component with homogeneous chemical composition and density, but an internal temperature inhomogeneity; the fact that in M 2-36 they don't agree has led us to consider the possibility that the different gas phases in M 2-36 may not be well mixed, arising the question of which $t^{2}$ is more representative of the photoionized region in a global scale, if any.
	
	\subsection{Chemical inhomogeneities}
	\label{subsec:chem_inhomogeneities}
	In the last twenty years, observational evidence for the pressence of chemical inhomogeneities has emerged. Notably, images of PN Abell 30 show clumps with ADFs exceeding 700 \citep{wesson03}, while the overall ADF of the object is around 70. Direct images of PNe have shown that emission from ORLs and CELs do not come from the same volume of gas \citep{garcia-rojas16}.

	As mentioned before, chemical inhomogeneities will produce temperature fluctuations, which can be modeled using the $t^{2}$ formalism. From a mathematical point of view however, the $t^{2}$ is limited to relatively small temperature fluctuations. Chemical inhomogeneities are necessary to reproduce large ADF values observed in photoionized regions. 
	
    The usual approach when modeling chemical inhomogeneities in 1D consists of incorporating several phases with different chemistry. Most commonly two-phase (bi-abundance) models are considered \citep[see][for example]{tsamis05, yuan11}. 
    
    Typically, two-phase models include a phase with "normal" chemical composition; and another one with high density, including either high metallicity or poor hydrogen content (the difference between both scenarios being the treatment of He; also high metallicity inclusions are usually helium poor).
	
    The implication of having a phase with very high density is that its mass becomes minimal, thus putting more weight on the hot phase, having a greater impact on the global chemistry of the object. 
    
    Consequentially, chemical inhomogeneities can be adapted to produce any ADF value.

    In the following section, we will show how a chemically inhomogeneous model reproduces important parameters reasonably.

	\section{A simple three-phase model}
	\label{sec:model}

	The differences between $T_{e}$(4363/4959), $T_{e}$(V1/4959), $T_{e}$(BJ) and $T_{e}$(He$^{+}$), are not consistent with a single phase model (not even one with an extreme $t^{2}$); in fact the wide dispersion of $t^{2}$ values present in Table \ref{tab:t2}, shows that these observations are not consistent with a chemically homogeneous PN. This in turn has led us to explore the possibility that more than one zone, each with radically different physical conditions, may coexist in the observed volume of PN M 2-36. 
	
    We have created ``simple" two- and three-phase models using the photoionization code Cloudy V.17 \citep{ferland17}; the PyCloudy library \citep{pycloudy} was also employed to write the Cloudy input files and define the weight of the gas phases. 
	
	The phases are distinguished mainly by their density and chemical composition (see Table \ref{tab:modelparam2}). In the two phase models we used a low-metallicity low-density  phase (Phase A), and a high metallicity phase; we explored the possibility of this second phase having a low to intermediate density, 10$^{3}$ $\le$ $n_{e}$ $\le$ 10$^{4}$ cm$^{-3}$ (we call this Phase B), or high density, $n_{e} \sim 10^{5}$ cm$^{-3}$  (we call this Phase C). For the three-phase models we included all three phases. The filling factor for each phase was also included as a variable in the model.

    Initially we considered a black body as an ionizing source, however, while we had some success reproducing the observed forbidden line ratios, many line intensities were unsatisfactory; notably, \ion{He}{ii} $\lambda$4686 was heavily overestimated. This led us to consider the PN atmosphere SEDs developed by \cite{rauch03} with solar metallicity.
    
    We find that a central star temperature ($T_{\it eff}$) of 80000 -- 90000 K is consistent with our observations \citep{gurzadian88}, hence we based our models around this interval. Since the conditions of the central star of M~2-36 are not well known, we sought a $T_{\it eff}$ value able to reproduce reasonably the observed intensities of \ion{He}{ii} $\lambda$4686 and \ion{C}{ii} $\lambda$4267.

	The final characteristic that is phase dependent is the relevance of each phase, which we weigh by a fraction of the total volume. Varying the weight of all three phases allows for too many degrees of freedom. Other characteristics of the model, such as radius, distance to the object, and the ratio of heavy elements to oxygen  are the same for all phases. The corresponding values are included in Table \ref{tab:modelparam1}.
	
	Our approach to construct a grid of models was as follows. Abundances were set based on the observational results and modified in steps of 0.1 dex. The density of each phase and the external radius were set so that hydrogen was fully ionized in all phases. Large changes in $log \enspace g$ mostly affected \ion{He}{ii} $\lambda$4686 negatively, hence it was set to 5.0 (using a larger value would require us to increase $T_{\it eff}$ above 90000 K). The best models were selected and refined in terms of line intensities and ratios, comparing these with the observational results. Different weights for each phase were explored in every iteration. 
	
    Overall we constructed three sets of models, one with phases A and B; one with phases A and C; and the last one with all tree phases. The best model from each set is presented in tables \ref{tab:modelparam2} (inputs) and \ref{tab:model_output} (outputs). Model 1 represents the best two-phased model that includes phases A and B, Model 2 represents the best two-phased model that includes phases A and C, and Model 3 represents the best three-phased model (we do not search for two phased models with phases B and C, because such high metallicities  will never be able to reproduce the observed CEL ratios). In order to have a meaningful ADF we need to combine phase A with phase B and/or phase C. 

    We selected our preferred models by trying to fit simultaneously the traditional forbidden line intensity ratios (those used to determine temperature and density), as well as the fraction of \ion{O}{ii} V1 attributed to \ion{O}{ii} $\lambda$4649 \citep[an ORL density indicator;][]{peimbert13}. The upper limit for this fraction is 0.40, which represents a global $n_{e}$ in excess of $10^{5}$ cm$^{-3}$, therefore we favored models that stay close to the observed value of 0.29, implying a density of 1100 cm$^{-3}$.
 
\begin{figure*}
    \centering
    \includegraphics[width=\textwidth]{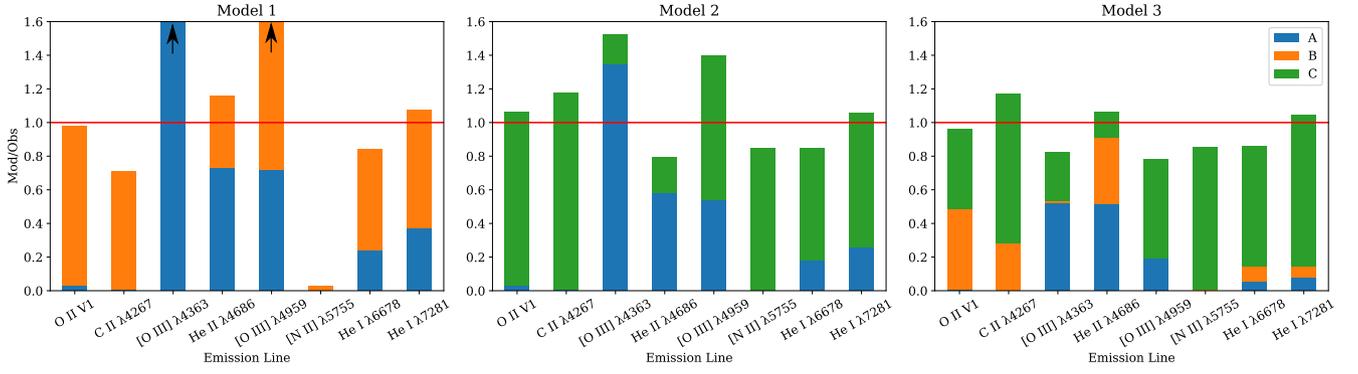}
    \caption{Comparison between model output and observations. Also shown is the contribution of each phase to line intensities.}
    \label{fig:barplot}
\end{figure*}

    We present the fractional contributions from each phase to our three preferred models in Figure \ref{fig:barplot}. Overall we can see that no phase dominates emission in any of the models. 
    
    In all models, Phase A is more evident in [\ion{O}{iii}] CELs and \ion{He}{ii} $\lambda$4686, with a small contribution to the \ion{He}{i} recombination lines, but being barely present in the heavy element recombination lines. High metallicity phases (either B, C, or both) dominate ORL emission in both two and three-phase models. 
    
    We chose to include the \ion{O}{ii} $\lambda$4649/V1 ratio, since it is one of a very small set of line diagnostics that focuses on the cold/high metallicity gas.  That being said, none of the two-phase models is able to simultaneously reproduce [\ion{O}{iii}] $\lambda$4959/$\lambda$4363 and \ion{O}{ii} $\lambda$4649/V1 ratios (while reproducing the observed contrasts between CELs and ORLs). Model 1  reproduces the total intensity of \ion{O}{ii} V1 successfully and the $\lambda$4649/V1 ratio acceptably, but fails to reproduce the CEL emission and ratios. The overall [\ion{O}{iii}] $\lambda$4959/$\lambda$4363 remains too low (too hot). In Model 2, the very high density of phase C lowers the cooling efficiency of the heavy elements allowing for moderate temperatures. However, the fraction of 4649/V1 reaches an extreme value that is irreconcilable with our observations implying a total density of $\sim10^{5}$ cm$^{-3}$. After running hundreds of two-phase models, these trends were unavoidable. A combination of phases B and C is necessary to reproduce the \ion{O}{ii} $\lambda$4649/V1 ratio in the high metallicity gas required by the observed ADF. 

    The idea of two-phased models has been proposed before to explain the abundances observed in some PNe and H II regions \citep{liu00, tsamis05, yuan11, danehkar18}, and has been expanded recently by \cite{gomezllanos20} (who constructed a bi-abundance, four-phased model for NGC 6153). Most of the two phased models in the literature include phases equivalent to our phases A and C, i.e. are equivalent to model 2; nonetheless we consider that it is important to try to place some constraints specific to the cold high metallicity gas, as we always do when working with the hot, low metallicity gas (such as the \ion{O}{ii} $\lambda$4649/V1 ratio). When such additional constraints are included the results from models 1 and 2 fall short of being satisfactory, and force us into using more than two phases when modeling M 2-36. 

    We ran hundreds of three-phase models searching for a better fit. The best three-phase models are much better suited to reproduce our constraints than the best two-phase models. We are now able to reproduce the observed [\ion{O}{iii}] $\lambda$4959/$\lambda$4363 ratio, the \ion{O}{ii} V1 emissivity and an important departure from equilibrium of the $\lambda$4649/V1 ratio.  Overall we are also satisfied with individual emission lines, and the other ratios presented in Table \ref{tab:model_output}. We simultaneously reproduce some of the brightest emission lines with a maximum deviation of 22\%. Still, a finer grid of models around these values is necessary to obtain better fits.

	In the presence of chemical inhomogeneities the abundances present in any phase (or element of volume) are not representative of the entire object; neither CEL abundances nor ORL abundances are truly representative of the object as a whole. \cite{gomezllanos20} argue that the Abundance Contrast Factor (ACF), defined as the ratio of metal-rich to "normal" components in photoionization models is a better value for determining the true abundance difference than the ADF. In Models 1, 2, and 3, the logarithmic value of ACF(O) is 1.3, 1.63 and 1.63 respectively, implying that ADF(O$^{++}$) from our models is overestimated. Indeed, ADF(O$^{++}$) obtained from our model outputs is 4.98, 6.66 and 10.16. Taking their work as reference, a more realistic ADF(O$^{++}$) would be around 1.5 for all three models.
	
	Observationally it is very difficult to properly characterize each of the phases (or even to determine the number of phases present in an object), and thus it is difficult to assess the global quantity of any element. On the other hand, when constructing a model we can have specific quantities of set phases built-in and thus have complete knowledge of the total quantities for all elements; in other words, we can "count" the atoms present in each phase and add them up. It must be emphasized that these abundances can not be measured directly from either real or synthetic spectra.
	
    The total He/H and O/H abundances included in each model can be seen in Table \ref{tab:abund_model}. In all cases the abundances of the models are somewhat intermediate between those from the low metallicity and high metallicity phases.  The O/H abundances are also intermediate between the observed CEL and ORL abundances; the exact value strongly depends on the mass of each phase, since the emission measure is somewhat similar for all phases, the lower the density of any given phase, the more relevant the chemical abundance of said phase is to the total object.
    
    The high metallicity phases of our models have higher abundances than those determined from ORL  (about 15\% for He and a about factor of 2 to 4 for O). This arises because there is a significant contribution to the H lines from the low metallicity phase, thus diluting the O/H abundance seeded in the high metallicity phases.
    
	In our interpretation of these three phases: phase C represents high-density, high-metallicity, low-volume inclusions which are probably being evaporated and thus are surrounded by a low-density, high-metallicity,  high-volume medium (phase B); all these are immersed in a low-density, low-metallicity, high-volume "normal" medium (phase A).  An interesting question is whether phase C represents nodules, filaments, or shells. This model does not really discriminate between those options, any of them can be represented by model 3.

    As a last note on photoionization models, we would be remiss not to mention the possibility of the existence of more than 3 phases, or more accurately, more than 3 phases with a significant contribution to the overall line emission. While it is indeed possible, current observations do not give us enough constraints to discern the unique characteristics of so many phases, and a photoionization model, even a toy model, has too many degrees of freedom and thus the parameter space to study becomes too large to properly explore. Altogether it is beyond the scope of this paper to try to explore any such model.

\begin{table}
\caption{Input parameters of the best-fit models}
\centering
\label{tab:modelparam1}
\begin{tabular}{l c}
\hline
Parameter & Value\\
\hline
\hline
Ionization Source & PN SED \citep{rauch03} \\
log $g$ (source) & 5.0 \\
Distance (pc) & 6100$^a$ \\
Ext. Radius (cm) & $10^{18.14}$ \\
log(C/O) & -0.53 \\
log(N/O) & -1.38 \\ 
log(Ne/O) & -0.98 \\
log(S/O) & -3.24 \\ 
log(Cl/O) & -3.1 \\
log(Ar/O) & -2.27 \\ \hline
\multicolumn{2}{l}{$^a$ \cite{stanghellini08}} \\
\end{tabular}
\end{table}
	
\begin{table*}
\caption{Parameters of the best-fit models.}
\centering
\label{tab:modelparam2}
\begin{tabular}{l c c c c c c c c } \hline
& \multicolumn{4}{c}{Two Phase Models}  & \multicolumn{3}{c}{Three Phase Model} \\ 
\cline{2-5} \cline{6-8~} 
& \multicolumn{2}{c}{1} & \multicolumn{2}{c}{2} & \multicolumn{3}{c}{3}  \\
\cline{2-3} \cline{4-5~} \cline{6-8~} 
Phase                   & A & B & A & C & A &	B & C  \\
\hline
$T_{\it eff}$ (K) & \multicolumn{2}{c}{90000} &	\multicolumn{2}{c}{82000} & \multicolumn{3}{c}{84000}  \\
$Q$(H$^{0}$) & \multicolumn{2}{c}{10$^{48.6}$} & \multicolumn{2}{c}{10$^{48.9}$} &\multicolumn{3}{c}{10$^{48.8}$}  \\
Density	(cm$^{-3}$)     &$10^{3}$& $10^{4}$  & 10$^{2.9}$& 10$^{5}$ & 10$^{2.7}$        & 10$^{3.0}$               & 10$^{5.3}$   \\
Filling factor          & 0.3   & 0.005        & 0.3    & 0.00005     &       0.25              & 0.011                 & 0.00001   \\
Weight (Emis. measure)  & $^3$\hspace{-1pt}$/_5$	 & $^2$\hspace{-1pt}$/_5$         & $^3$\hspace{-1pt}$/_5$     &$^2$\hspace{-1pt}$/_5$       & $^1$\hspace{-1pt}$/_3$	& $^1$\hspace{-1pt}$/_3$& $^1$\hspace{-1pt}$/_3$   \\ 
log(He/H)               & -0.880 & -0.834 &  -0.880 & -0.834 &    -0.880	& -0.834  &   -0.834 \\
log(O/H)                & -3.35    & -2.05  &  -3.35  & -1.72   & -3.35 & -1.72 & -1.72 	\\
\hline
\end{tabular}
\end{table*}	

\begin{table*}
\caption{Line intensities outputs of the best-fit models.}
\label{tab:modelparam_out}
\centering
\begin{tabular}{l c cc cc cc} \hline
  & \multicolumn{7}{c}{$I(\lambda$)/$I({\rm H}\beta)$} \\ \hline
  & & \multicolumn{2}{c}{Model 1} &\multicolumn{2}{c}{Model 2} &\multicolumn{2}{c}{Model 3}   \\
  \cline{3-4} \cline{5-6~} \cline{7-8~} \\
Emiss. line  & Obs. & Mod & M/O & Mod & M/O & Mod &  M/O \\ \hline 
\ion{O}{ii} V1 sum.                          & 4.84   & 4.73   & 0.98 & 5.14   & 1.06 &   4.63  & 0.96     \\ 
{[\ion{O}{ii}]} $\lambda\lambda$3726$+$29    &  1.84  & 90.85  & 49.38 & 2.55   & 1.38 &   2.61  & 1.42        \\
{\ion{C}{ii}} $\lambda$4267                & 2.34   &  1.67  & 0.71 & 2.76    & 1.18       & 2.74 & 1.17 \\ 
{[\ion{O}{iii}]}$\lambda$4363                & 2.8    & 5.54   & 1.98 & 4.26   & 1.52 &   2.28	& 0.81     \\
\ion{He}{ii} $\lambda$4686                   & 4.122  & 4.78   & 1.16 & 3.27   & 0.79 &   4.37	& 1.06    \\
\null [\ion{O}{iii}] $\lambda$4959           & 263.64 & 458.79 & 1.74 & 367.82 & 1.40 &   206.43& 0.78     \\
{[\ion{N}{ii}]} $\lambda$5755              & 4.907 & 0.13    & 0.03 & 4.14   & 0.84 &   4.14	& 0.84         \\
\ion{He}{i} $\lambda$6678                    & 5.88  & 4.96    & 0.84 & 5.01   & 0.85 &   5.07	& 0.86            \\
\ion{He}{i} $\lambda$7281                    & 0.86  & 0.92    & 1.07 & 0.91   & 1.06 &   0.9	& 1.05     \\ \hline
Parameter &  &  & & & &  &   \\ \hline
\ion{O}{ii} $\lambda4649$/$\text{V1}$                     & 0.29  & 0.36  & 1.24 & 0.38  & 1.31 & 0.33  & 1.14   \\
{[\ion{O}{iii}]} $\lambda$4959/$\lambda$4363                 & 90.90 & 85.10 & 0.94 & 85.10 & 0.94 & 90.73 & 0.99  \\
{[\ion{N}{ii}]} $\lambda$6548+$\lambda$6584/$\lambda$5755    & 38.46 & 395.15 & 10.27   &   90.91   & 2.36      & 52.83 & 1.37 \\
{[\ion{O}{iii}]} $\lambda$4959/{[\ion{O}{ii}]} $\lambda\lambda$3726+29 & 4.16 & 5.05 & 1.20     &  3.29   & 0.78 & 2.60& 0.63  \\ \hline
\end{tabular}
\label{tab:model_output}
\end{table*}

\begin{table*}
    \caption{Chemical abundances from photoionization models}
    \centering
    \begin{tabular}{c c c c c c} \hline
        & \multicolumn{2}{c}{Obs} & \multicolumn{3}{c}{Models} \\
             \cline{2-3}  \cline{4-6~} \null \vspace{-7pt}\\
    Ratio  &  CEL  & ORL & M1 & M2 & M3  \\ \hline
    log(He/H) &  & -0.888 & -0.879 &  -0.879 &  -0.879 \\
    log(O/H) & -3.26  & -2.36 & -2.89 &  -3.15  &   -2.66 \\ \hline
    \end{tabular}
    \label{tab:abund_model}
\end{table*}

		\section{Summary and Conclusions}
		\label{sec:conclusions}
		
    We have analyzed a high resolution spectrum of the planetary nebula M 2-36 obtained with the Ultraviolet and Visual Echelle Spectrograph at the Very Large Telescope covering the spectral range 3030--10360 \AA. We measure the intensities of 420 emission features and, once blends are considered, we identify 446 emission lines. Notably, we identify three emission features from ions produced by the s-process ([\ion{Kr}{iv}] $\lambda$5346 and $\lambda$5868; and  [\ion{Xe}{iv}] $\lambda$7536) marking the first detection of said elements in this nebula. Consequently we compute the total abundance of Kr and Xe.
	
	We have computed $T_{e}$ and $n_{e}$ using traditional plasma diagnostics (from CELs). Our results are mostly consistent with a previous analysis of this object by \citet{liu01}. We also computed $T_{e}$ and $n_{e}$ from ORLs of He$^{+}$ and O$^{++}$ as well as from the Balmer jump (H$^{+}$). All our CEL determinations are consistent with each other within a two zone photoionization scheme; this is particularly significant for the temperatures, due to the small error bars of our determinations. 

	Chemical abundances are reported for a large sample of ions. We have computed ionic abundances for C$^{++}$, N$^{++}$, O$^{++}$, and Ne$^{++}$ from ORLs and have obtained the corresponding ADFs. Given the quality of our spectrum, we produce a robust result for ${\rm ADF(O}^{++})=6.76\pm0.50$, this value is slightly larger than the one measured by \cite{liu01}, but a difference is to be expected, since we are not observing the whole nebula (i.e. we do not observe the exact same volume). 
	 
	The large value of ADF(O$^{++}$) cannot be attributed solely to the presence of temperature variations in a chemically homogeneous PN, thus suggesting the presence of at least two volumes of gas with different chemical composition. 

	Using the $t^{2}$ formalism of \cite{peimbert67} we are able to determine three independent measurements for temperature inhomogeneities: $t^{2}$(O$^{++}$), $t^{2}$(He$^{+}$) and $t^{2}$(H$^{+}$). These determinations are not reconcilable under the assumption of chemical homogeneity; in fact the disparity between these three values reveals a complicated structure within the nebula and suggests the presence of more than two phases within the PN.

	Our $t^2$ determinations motivated us to search for a simple photoionization model able to reproduce the general characteristics of M 2-36. We explored the possibility that a chemically inhomogeneous two-phased model was able to reproduce the observed characteristics of M 2-36, but were generally underwhelmed by the results. This setback, and previous suggestions to the existence of more than two phases, led us to explore the possibility that chemically inhomogenous three-phase Cloudy models were better able to reproduce the observed emission of M 2-36. 
	
	Our favorite model is able to reproduce the most relevant observed quantities related to O$^{++}$ within an acceptable range, notably the nebular to auroral line  ratio, the total intensity of multiplet V1, and the ratio of the intensity of multiplet V1 of \ion{O}{II} to $I$(\ion{O}{ii} $\lambda$4649), as well as to reproduce the observed He$^+$ and He$^{++}$ line intensities.
	
	While we are well pleased with the way our model fits the observations, this does not mean that M 2-36 is only composed by large quantities of matter that can be aligned with these three phases. First, it is very likely that there are better three-phase models than the one we found; second, there could be an important fraction of the emission of M 2-36 that comes from matter that corresponds to additional phases.

	We must remark that the models presented in section \ref{sec:model} are only toy models. Our exploration has clear limitations: the grid we used could be refined, we do not explore in depth the abundances of other elements, it does not fully explore the effect of the mixing of three phases (it is the sum of three 1-D simulations), etc. 

     Our models show that, for M 2-36, the high metallicity gas must have at least some phase with low density (similar to that of the low metallicity gas). This means that there is a significant mass in the high metallicity gas, that must be taken into account when calculating the total abundances, and that the CEL abundances are not representative of the whole object (the fact that model 2 doesn't work means that we can not use the ORL abundances either). In fact our models show that the total abundances of M 2-36 are intermediate between CEL and ORL abundances.

    Unfortunately our search for the best fit also showed us that the precise determination of total abundances depends strongly on the ratios between the emission measure of phases B and C and of the exact density of phase B. 
    
    The complexity of fitting a three phase model with only one set of observations limits the relevance of the abundance determinations of any such model model. However, in this case, the exact O/H abundance appears to be close to the average between the CEL and ORL determinations.

	While our tolerances do not permit us to explore the possibility of finding a model with more than three phases that produces a significantly better fit to the observed line intensities, we must ask ourselves whether such model exists. Furthermore, would more than three phases be required to model the gas component if we had deeper observations?. More importantly, is this model adequate to represent the real physical conditions within M 2-36?. Answers to those questions are beyond the scope of this paper. 

    Despite having been studied previously, our deep spectral analysis allowed us to discover that the chemistry of M 2-36 may be more complicated than previously thought. Our observations allowed us to measure critical quantities that characterize the object revealing the presence of at least three phases in the gas. This was possible largely thanks to the availability of temperature diagnostics from ORLs, whose use is not widespread and which are not always available.  
   
    This highlights the necessity of analyzing more objects similar to M 2-36 (with high but not extreme ADF) using a similar methodology. Since we do not expect all other high-ADF PNe to behave in the same way, we propose that many other high ADF objects should be studied specifically looking for evidence of both: low-density and high-density in their high metallicity components, before arriving to any conclusion regarding this type of objects.
   
   In any case, it is clear that when determining the total abundances for PNe where the \ion{O}{ii} $\lambda$4649/VI ratio is less than $\sim 0.40$ the high metallicity regions can not be pushed aside and must be mixed with the low metallicity regions (i.e. when \ion{O}{ii} $\lambda$4649/VI is not consistent with 0.40 the CEL abundances should not automatically be considered the correct abundances).
    
    Regarding PNe with low to moderate ADFs, they should have several deep spectral analyses (independent of those of high ADF PNe) before arriving to any conclusion regarding these kind of objects and not to automatically assume objects like M 2-36 or Abell 30 can be used as models for objects with low ADF.

	\section*{Acknowledgements}
	
	We wish to thank Mar\'ia Teresa Ruiz and Manuel Peimbert for their invaluable help with the observations. We thank an anonymous referee for numerous comments and suggestions that helped us improve the paper. J. N. Esp\'iritu acknowledges the support of CONACyT grant 464709 and UNAM DGAPA/PAPIIT IG 100319 and IN103820. We also thank Ver\'onica G\'omez Llanos for her help with Python and PyCloudy. The manuscript of this paper has been written in the Overleaf environment.

	\begin{table*}
		\caption{Emission line list}
		\label{tab:linelist}
		\begin{tabular}{c c c c c c c c}
			\hline
			$\lambda_{0}$ & $\lambda_{\text{obs}}$ & Ion & Identification & F & I & Error
			(\%) & Notes \\
			\hline
			\hline
			3187.74 	& 	3188.20 	&	 \ion{He}{I} 	&  	3 			& 1.652 		&  	2.578 	& 	8 	&\\
			3203.10	&	3203.57	&	\ion{He}{II} 	& 	 3			& 	0.542	&	0.817	&	13 	&\\
			3218.19	&	3218.64	&	\ion{Ne}{II}	&	4D$_{0}$--4F	&	0.177	&	0.265	&	23 	&\\
			3244.00	&	3244.54	&	\ion{O}{II}		&	4P--D[2]$_{0}$	&	0.086	&	0.127	&	32 	&\\
			3334.87	&	3335.31	&	\ion{Ne}{II}	&	2			&	0.390	&	0.559	&	15	&\\
			3354.56	&	3355.01	&	\ion{He}{II}	&	1S--1P$_{0}$	&	73.889	&	105.194	&	2	&\\
			3367.05	&	3367.70	&	\ion{Ne}{II}	&	12			&	0.106	& 	0.150	&	29	&  \\
			3444.02	&	3444.55	&	\ion{He}{I}		&	3P$_{0}$--3S	&	1.327	&	1.845	&	8	& ? \\
			3447.63	&	3448.08	&	\ion{He}{I}		&	3P$_{0}$--3D	&	0.132	&	0.183	&	26	& \\
			3487.72	&	3488.13	&	\ion{He}{I}		&	3P$_{0}$--3D	&	0.068	&	0.093	&	36	& \\ 
			3498.64	&	3499.09	&	\ion{He}{I}		&	3P$_{0}$--3D	&	0.093	&	0.128	&	31	& \\
			3512.51	&	3513.02	&	\ion{He}{I} 	&	3P$_{0}$--3D	&	0.138	&	0.189	&	26	& \\	
			3530.50	&	3530.97	&	\ion{He}{I} 	&	3P$_{0}$--3D	&	0.139	&	0.190	&	26	& \\
			3554.42	&	3554.88	&	\ion{He}{I} 	&	3P$_{0}$--3D	&	0.196	&	0.266	&	22	& \\
			3587.28	&	3587.79	&	\ion{He}{I}		&	3P$_{0}$--3D	&	0.295	&	0.399	&	18	& \\
			3613.64	&	3614.15	&	\ion{He}{I}		&	3P$_{0}$--3D	&	0.277	&	0.372	&	18	& \\
			3634.25	&	3634.75	&	\ion{He}{I}		&	3P$_{0}$--3D	&	0.434	&	0.582	&	14	& \\
			3656.56	&	3657.16	&	\ion{H}{I}				&	H37	&	0.143	&	0.190	&	25 & \\
			3657.27	&	3657.77	&	\ion{H}{I}				&	H36	&	0.143	&	0.190	&	25 & \\
			3657.92	&	3658.41	&	\ion{H}{I}				&	H35	&	0.138	&	0.184	&	26 & \\
			3658.64	&	3659.11	&	\ion{H}{I}				&	H34	&	0.156	&	0.208	&	24 & \\
			3659.42	&	3659.95	&	\ion{H}{I}				&	H33	&	0.153	&	0.203	&	24 &  \\
			3660.28	&	3660.79	&	\ion{H}{I}				&	H32	&	0.174	&	0.232	&	23 & \\
			3661.22	&	3661.71	&	\ion{H}{I}				&	H31	&	0.205	&	0.273	&	21 &  \\
			3662.26	&	3662.75	&	\ion{H}{I}				&	H30	&	0.229	&	0.305	&	20 & \\
			3663.40	&	3663.90	&	\ion{H}{I}				&	H29	&	0.230	&	0.306	&	20 &  \\
			3664.68	&	3665.14	&	\ion{H}{I}				&	H28	&	0.280	&	0.373	&	18 & \\ 
			3666.10	&	3666.61	&	\ion{H}{I}				&	H27	&	0.299	&	0.398	&	17 & \\
			3667.68	&	3668.18	&	\ion{H}{I}				&	H26	&	0.299	&	0.398	&	17 & \\
			3669.46	&	3669.97	&	\ion{H}{I}				&	H25	&	0.322	&	0.428	&	17 & \\
			3671.48	&	3671.99	&	\ion{H}{I}				&	H 24	&	0.364	&	0.485	&	16 & \\
			3673.76	&	3674.28	&	\ion{H}{I}				&	H23	&	0.361	&	0.480	&	16 & \\
			3676.36	&	3676.36	&	\ion{H}{I}				&	 H22	&	0.378	&	0.503	&	16 & \\
			3679.35	&	3679.87	&	\ion{H}{I}				&	H 21 &	0.432	&	0.574	&	14 & \\
			3682.81	&	3683.32	&	\ion{H}{I}				&	H20	&	0.516	&	0.685	&	13 & \\
			3686.83	&	3687.34	&	\ion{H}{I}				&	H19  &	0.598	&	0.794	&	12 & \\
			3691.55	&	3692.07	&	\ion{H}{I}				&	H 18	&	0.750	&	0.995	&	11 & \\
			3694.22	&	3694.73	&	\ion{Ne}{II}			&	V1	&	0.270	&	0.358	&	18 & \\
			3697.15	&	3697.66	&	\ion{H}{I}				&	H 17	&	0.841	&	1.115	&	10 & \\
			3702.62	&	3703.22	&	\ion{He}{I}	            &	1P$_{0}$-1D&	0.088	&	0.117	&	32 &  \\
			3703.86	&	3704.36	&	\ion{H}{I}				&	H 16	&	0.947	&	1.254	&	10 & \\
			3705.02	&	3705.52	&	\ion{He}{I}				&	3P$_{0}$-3D	&	0.719	&	0.951	&	11 & \\
			3707.25	&	3707.76	&	\ion{He}{I}				&	1P$_{0}$-1S	&	0.081	&	0.107	&	33 & \\
			3709.62	&	3710.12	&	\ion{Ne}{II}			&	V1		&	0.069	&	0.091	&	36 & \\
			3711.97	&	3712.49	&	\ion{H}{I} 				&	H 15		&	1.158	&	1.530	&	9 & \\
			3713.08	&	3713.54	&	\ion{Ne}{II}			&	V5		&	0.262	&	0.346	&	19 & \\
			3721.87	&	3722.38	&	[\ion{S}{III}] 	        &	2F 	&	2.072	&	2.733	& 7 & \\
			3721.93 &           &   \ion{H}{I}              &    H 14    &          &           &   & \\
			3726.03	&	3726.57	&	[\ion{O}{II}] 			&	1F 	&	31.155	&	41.072	&	2 & \\
			3727.25	&	3727.80	&	\ion{N}{II}				&	3P$_{0}$-3D		&	0.325	&	0.429	&	17 & \\
			3728.82	&	3729.32	&	[\ion{O}{II}] 		    &	1F		&	16.938	&	22.319	&	3 & \\
			3734.37	&	3734.89	&	\ion{H}{I}			    &	H13		&	1.771	&	2.332	&	7 & \\
			3750.15	&	3750.65	&	\ion{H}{I} 			       &	H12		&	2.370	&	3.111	&	6 & \\
			3754.69	&	3755.22	&	\ion{O}{III}			    &	V2		&	0.151	&	0.199	&	24 & \\
			3757.24	&	3757.75	&	\ion{O}{III}	            &	V2	&	0.049	&	0.065	&	28 & \\
			3759.87	&	3760.41	&	\ion{O}{III}                &	V2	&	0.337	&	0.442	&	11 & \\
			3762.47	&	3762.99	&	\ion{O}{II} 				&	V31	&	0.039	&	0.052	&	32 & \\
			3766.26	&	3766.78	&	\ion{Ne}{II}				&	V1			&	0.060	&	0.078	&	26 & \\
			3770.63	&	3771.15	&	\ion{H}{I}					&	H11			&	2.704	&	3.536	&	4  & \\
			3774.02	&	3774.54	&	\ion{O}{III}				&	V2			&	0.047	&	0.062	&	29 & \\
			3777.14	&	3777.67	&	\ion{Ne}{II}				&	V1		&	0.071	&	0.093	&	24 & \\
			3784.89	&	3785.43	&	\ion{He}{I}					&	V64		&	0.033	&	0.042	&	35 & \\
			3791.27	&	3791.84	&	\ion{O}{III}				&	V2		&	0.051	&	0.066	&	28 & \\
		\end{tabular}
	\end{table*}
	
	\begin{table*}
		\contcaption{}
		\label{tab:continued1}
		\begin{tabular}{c c c c c c c c}
			\hline
			$\lambda_{0}$ & $\lambda_{\text{obs}}$ & Ion & Identification & F & I & Error
			(\%) & Notes \\
			\hline
			\hline
			3797.90	&	3798.43	&	\ion{H}{I}                 &	H10		&	3.658	&	4.759		&	4 & \\
			3805.74	&	3806.31	&	\ion{He}{I}					&	V63			&	0.070	&	0.091	&	24 & \\
			3806.49	&	3807.01	&	\ion{Si}{III}				&	V5			&	0.041	&	0.054	&	31 & \\
			3819.61	&	3820.17	&	\ion{He}{I} 				&	V22			&	1.261	&	1.634	&	6 & \\
			3829.77	&	3830.25	&	\ion{Ne}{II}				&	V39			&	0.037	&	0.048	&	33 & \\
			3833.55	&	3834.08	&	\ion{He}{I}					&	V62	    &	0.066	&	0.086	&	24 & \\
			3835.39	&	3835.92	&	\ion{H}{I}					&	H 9			&	5.409	&	6.986	&	3 & \\
			3856.02	&	3856.58	&	\ion{Si}{II}				&	V1			&	0.115	&	0.147	&	19 & \\
			3862.59	&	3863.14	&	\ion{Si}{II}				&	V1			&	0.135	&	0.174	&	17 & \\
			3868.75	&	3869.30	&	[\ion{Ne}{III}]				&	F1			&	62.759	&	80.525	&	2 & \\
			3871.79	&	3872.32	&	\ion{He}{I}					&	1P$_{0}$-1D	&	0.084	&	0.108	&	22 & \\
			3880.33	&	3880.80	&	\ion{Ar}{II}				&	54			&	0.046	&	0.059	&	29 & \\
			3882.19	&	3882.76	&	\ion{O}{II}					&	V12			&	0.116	&	0.149	&	18 & \\
		    3888.65	&	3889.42	&	\ion{He}{I}	             	&	V2   		&	16.892	& 21.584    &	2 & \\
		    3889.05 &           &   \ion{H}{I}                  &   H8          &           &           &   & \\
			3912.72	&	3913.32	&	\ion{Mn}{I}					&				&	0.037	&	0.047	&	33 & ? \\
			3920.59	&	3921.18	&	\ion{C}{II}					&	V4			&	0.040	&	0.051	&	31 & \\
			3926.53	&	3927.11	&	\ion{He}{I}					&   V58           &	0.125	&	0.159	&	18 & \\
			3964.74	&	3965.28	&	\ion{He}{I}		            &	V5	&	0.891	&	1.120	&	7	&	\\
			3967.46	&	3967.98	&	[\ion{Ne}{III}] 	&	1F			&	11.000	&	13.819	&	2	&	\\
			3968.43 &           &   \ion{He}{I}                 &       4.14        &           &           &       &   \\
			3970.07	&	3970.63	&	\ion{H}{I}		&	H7			&	15.857	&	19.910	&	2	&	\\
			3973.24	&	3973.81	&	\ion{O}{II}		&	6			&	0.103	&	0.130	&	20	&	\\
			3979.78	&	3980.29	&	[\ion{Fe}{II}]  	&	F9			&	0.049	&	0.061	&	28	&	\\
			3979.93	&			&	[\ion{Fe}{II}]	&	F8			&			&			&		&	\\
			3994.99	&	3995.54	&	\ion{N}{II}		&	V12			&	0.050	&	0.063	&	28	&	\\
			4009.26	&	4009.82	&	\ion{He}{I}		&	V55	        &	0.269	&	0.335	&	12	&	\\
			4013.99	&	4014.59	&	\ion{Ne}{I}		&	2			&	0.033	&	0.041	&	35	&	\\
			4023.98	&	4024.51	&	\ion{He}{I}		&	1P$_{0}$-1S	&	0.032	&	0.040	&	35	&	\\
			4026.21	&	4026.76	&	\ion{He}{I}		&	V18	&	4.649	&	5.762	&	3	&	\\
			4035.07	&	4035.65	&	\ion{O}{II} 	&	V68			&	0.113	&	0.139	&	19	&	\\
			4035.08	&			&	\ion{N}{II}		&	V39a			&			&			&		&	\\
			4041.31	&	4041.88	&	\ion{N}{II}		&	V39			&	0.201		&	0.248	&	14	&	\\
			4043.53	&	4044.08	&	\ion{N}{II}		&	V39			&	0.074		&	0.091	&	23	&	\\
			4062.94	&	4063.51	&	\ion{O}{II}		&	V50	        &	0.131		&	0.161	&	17	&	\\
			4068.60	&	4069.14	&	[\ion{S}{II}]  	&	1F 		    &	4.339	&	5.323	&	3	&	\\
			4069.62	&	4070.38	&	\ion{O}{II}		&	V10 			&	1.412	&	1.732	&	5	&	\\
			4069.89	&			&	\ion{O}{II}		&	V10			&			&			&		&	\\
			4071.24	&	4071.80	&	\ion{O}{II}	    &	V48a	    &	0.084	&	0.103	&	22	&	\\
			4072.15	&	4072.73	&	\ion{O}{II}		&	V10	        &	0.894		&	1.096	&	7	&	\\
			4075.86	&	4076.51	&	\ion{O}{II} 	&	V10 		    &	1.368	&	1.676	&	6	&	\\
			4076.56	&	4077.12	&	\ion{S}{II}	    &	F1			&	0.519	&	0.635	&	9	&	\\
			4078.84	&	4079.41	&	\ion{O}{II}	    &	V10	        &	0.117		&	0.144	&	18	&	\\
			4083.89	&	4084.46	&	\ion{O}{II}	    &	V48b 		&	0.123	&	0.151	&	18	&	\\
			4085.11	&	4085.68	&	\ion{O}{II}	    &	V10	        &	0.146		&	0.178	&	16	&	\\
			4087.15	&	4087.72	&	\ion{O}{II}	    &	V48	        &	0.143		&	0.174	&	17	&	\\
			4089.29	&	4089.84	&	\ion{O}{II}	    &	V48	        &	0.469		&	0.572	&	9	&	\\
			4092.93	&	4093.50	&	\ion{O}{II}	    &	V10	        &	0.110		&	0.135	&	19	&	\\
			4095.66	&	4096.23	&	\ion{O}{II}	    &	V48	        &	0.113		&	0.138	&	19	&	\\
			4097.32	&	4097.88	&	\ion{O}{II} 	&	V48 	    &	1.856	&	2.261	&	5	&	\\
			4098.24	&	4098.79	&	\ion{O}{II}	    &	V46a	    &	0.110	&	0.134	&	19	&	\\
			4100.05	&	4100.62	&	\ion{He}{II}	&	V4	        &	0.113		&	0.138	&	19	&	\\
			4101.75	&	4102.32	&	\ion{H}{I}  	&	H 6	        &	33.631		&	40.926	&	2	&	\\
			4103.39	&	4103.95	&	\ion{N}{III}	&	V1	        &	1.010		&	1.229	&	6	&	\\
			4104.95	&	4105.51	&	\ion{O}{II}	    &	V20	        &	0.192		&	0.233	&	14	&	\\
			4107.10	&	4107.67	&	\ion{O}{II} 	&	V48			&	0.098	&	0.119	&	20	&	\\
			4110.79	&	4111.37	&	\ion{O}{II}	    &	V20	        &	0.094		&	0.115	&	20	&	\\
			4119.22	&	4119.79	&	\ion{O}{II}	    &	V20	        &	0.270		&	0.328	&	12	&	\\
			4120.82	&	4121.31	&	\ion{He}{I}	    &	V16          &	0.316		&	0.383	&	11	&	\\
			4121.46	&	4122.07	&	\ion{O}{II} 	&	V19		    &	0.069	&	0.083	&	24	&	\\
			4132.80	&	4133.37	&	\ion{O}{II}	    &	V19	&	0.138		&	0.167	&	17	&	\\
			4143.76	&	4144.33	&	\ion{He}{I}	    &	V53	&	0.494		&	0.595	&	9	&	\\
			4145.90	&	4146.66	&	\ion{O}{II}	    &	V106			&	0.040	&	0.048	&	31	& \\
		\end{tabular}
	\end{table*}
	
	\begin{table*}
		\contcaption{}
		\label{tab:continued2}
		\begin{tabular}{c c c c c c c c}
			\hline
			$\lambda_{0}$ & $\lambda_{\text{obs}}$ & Ion & Identification & F & I & Error
			(\%) & Notes \\
			\hline
			\hline
			4146.08	&			&	\ion{O}{II}	    &	V106 			&			&			&		&	\\
			4153.30	&	4153.87	&	\ion{O}{II}	    &	V19	&	0.268		&	0.322	&	12	&	\\
			4156.53	&	4156.96	&	\ion{O}{II}	    &	V19			&	0.116	&	0.139	&	18	&	\\
			4168.97	&	4169.73	&	\ion{He}{I}     &	V52		&	0.119	&	0.143	&	18	&	\\
			4171.61	&	4172.20	&	\ion{N}{II}	    &	V43a	&	0.036	&	0.043	&	33	&	\\
			4176.16	&	4176.71	&	\ion{N}{II}	    &	V43a	&	0.066		&	0.079	&	24	&	\\
			4185.45	&	4186.01	&	\ion{O}{II} 	&	V36	&	0.089		&	0.106	&	21	&	\\
			4186.89	&	4187.47	&	\ion{C}{III} 	&	V18	&	0.061		&	0.073	&	25	&	\\
			4189.78	&	4190.36	&	\ion{O}{II}	    &	V36	&	0.114		&	0.135	&	19	&	\\
			4195.76	&	4196.33	&	\ion{N}{III}   	&	V6			&	0.041	&	0.048	&	31	&	\\
			4199.83	&	4200.56	&	\ion{He}{II}	&	4.11	&	0.137		&	0.162	&	17	&	\\
			4217.15	&	4217.73	&	\ion{N}{II}	&	3Po -- 3P				&	0.034	&	0.040	&	34	&	\\
			4219.76	&	4220.32	&	\ion{Ne}{II}	&	V52	&	0.097		&	0.115	&	20	&	\\
			4231.53	&	4232.18	&	\ion{Ne}{II}	&	V52b		&	0.039	&	0.046	&	32	&	\\
			4236.91	&	4237.56	&	\ion{N}{II} 	&	 V48			&	0.106	&	0.125	&	19	&	\\
			4237.05	&			&	\ion{N}{II}	&	V48			&			&			&		&	\\
			4241.78	&	4242.38	&	\ion{N}{II}	&	V48			&	0.163	&	0.192	&	16	&	\\
			4254.00	&	4254.61	&	\ion{O}{II} &	109			&	0.043	&	0.050	&	30	&	\\
			4267.15	&	4267.76	&	\ion{C}{II}	&	V6	&	2.011	&	2.344	&	5	&	\\
			4275.55	&	4276.16	&	\ion{O}{II}	&	V67		&	0.222	&	0.258	&	13	&	\\
			4276.75	&	4277.26	&	\ion{O}{II}	&	V67		&	0.095	&	0.111	&	20	&	\\
			4277.43	&	4278.02	&	\ion{O}{II} &	V67c	&	0.050	&	0.058	&		28	&	\\
			4281.32	&	4281.92	&	\ion{O}{II} &	53b		&	0.018	&	0.021	&	46	&	\\
			4282.96	&	4283.58	&	\ion{O}{II} &	67c		&	0.061	&	0.070	&	25	&	\\
			4283.73	&	4284.32	&	\ion{O}{II}	&	V67c	&	0.032	&	0.038	&	35	&	\\
			4285.70	&	4286.29	&	\ion{O}{II}	&	V78b	&	0.070	&	0.081	&	24	&	\\
			4288.81	&	4289.40	&	\ion{O}{II}	&	V53c	&	0.026	&	0.030	&	39	&	\\
			4291.25	&	4291.86	&	\ion{O}{II}	&	V55	    &	0.076	&	0.088	&	23	&	\\
			4292.21	&	4292.85	&	\ion{O}{II}	&	V78c	&	0.061	&	0.071	&	25	&	\\
			4294.92	&	4295.40	&	\ion{O}{II}	&	V54	&	0.125	&	0.145	&	18	&	\\
			4303.61	&	4304.40	&	\ion{O}{II} &	V65a 	&	0.238	&	0.275	&	13	&	\\
			4303.82	&			&	\ion{O}{II}	&	 V53a &			&			&		&	\\	
			4307.23	&	4307.82	&	\ion{O}{II} &	V54		&	0.044	&	0.051	&	30	&	\\
			4309.02	&	4309.61	&	\ion{O}{II} &	4D -- D[1]$_{0}$	&	0.029	&	0.033	&	37	&	\\
			4313.44	&	4314.03	&	\ion{O}{II}	&	V78a	&	0.042	&	0.048	&	31	&	\\
			4315.40	&	4316.00	&	\ion{O}{II} 	&	V63c &	0.031	&	0.035	&	36	&	\\
			4317.15	&	4317.74	&	\ion{O}{II}		&	V2	&	0.152	&	0.175	&	16	&	\\
			4319.63	&	4320.22	&	\ion{O}{II}		&	V2	&	0.106	&	0.122	&	19	&	\\
			4325.76	&	4326.36	&	\ion{O}{II}		&	V2	&	0.037	&	0.043	&	32	&	\\
			4329.75	&	4330.37	&	\ion{C}{II}		&	2D--Fo	&	0.013	&	0.015	&	54	&	\\
			4331.14	&	4331.74	&	\ion{O}{II}	    &	V65b	&	0.036	&	0.041	& 33	&	\\
			4332.71	&	4333.32	&	\ion{O}{II}		&	65		&	0.058	&	0.066	&	26	&	\\
			4336.86	&	4337.46	&	\ion{O}{II}		&	V2	&	0.053	&	0.061	&	27	&	\\
			4338.69	&	4339.29	&	\ion{He}{II}	&	4--10	&	0.063	&	0.072	&	25	&	\\
			4340.47	&	4341.07	&	\ion{H}{I}		&	H5 	&	39.645	&	45.317	&	2	&	\\
			4344.39	&	4344.99	&   \ion{O}{I}] 	&		&	0.038	&	0.044	&	32	&	?\\
			4345.56	&	4346.16	&	\ion{O}{II}	    &	V2	&	0.175	&	0.199	&	15	&	\\
			4349.43	&	4350.03	&	\ion{O}{II}	    &	V2	&	0.310	&	0.353	&	11	&	\\
			4351.51	&	4352.11	&	\ion{O}{II}	    &	V16	&	0.065	&	0.074	&	25	&	\\
			4353.60	&	4354.20	&	\ion{O}{II}	    &	V76c	&	0.046	&	0.052	&	29	&	\\
			4357.27	&	4357.87	&	\ion{O}{II}	    &	V63a	&	0.027	&	0.031	&	38	&	\\
			4363.21	&	4363.81	&	[\ion{O}{III}]	&	2F	&	2.465	&	2.800	&	4	&	\\
			4366.89	&	4367.47	&	\ion{O}{II} 	&	V2 	&	0.185	&	0.210	&	15	&	\\
			4371.59	&	4372.21	&	\ion{O}{II}		&	V76b	&	0.055	&	0.063	&	27	&	\\
			4379.11	&	4379.76	&	\ion{N}{III}	&	V18b	&	0.380	&	0.430	&	10	&	\\
			4379.55 &           &   \ion{Ne}{III}   &   V60b    &           &           &       &   \\
			4387.93	&	4388.54	&	\ion{He}{I}			&	V51	&	0.694	&	0.783	&	8	&	\\
			4391.94	&	4392.60	&	\ion{Ne}{II}		&	V55e	&	0.118	&	0.133	&	18	&	\\
			4397.98	&	4398.58	&	\ion{Ne}{ii}	&	V57b	&	0.027	&	0.031	&	38	&	\\
			4409.30	&	4409.91	&	\ion{Ne}{II}	&	V55e	&	0.094	&	0.106	&	20	&	\\
			4413.11 &   4413.81 &   \ion{Ne}{II}    & V57c          & 0.036 &	0.041  & 33 & \\
			4413.11 &           &    \ion{Ne}{II}   & V65           &       &           & \\
			4413.22	&	    	&	\ion{Ne}{II}    &	V55 &	 &		&		&	\\
		\end{tabular}
	\end{table*}
	
	\begin{table*}
		\contcaption{}
		\label{tab:continued3}
		\begin{tabular}{c c c c c c c c}
			\hline
			$\lambda_{0}$ & $\lambda_{\text{obs}}$ & Ion & Identification & F & I & Error
			(\%) & Notes \\
			\hline
			\hline
			4414.90	&	4415.51	&	\ion{O}{II}			&	 V5	&	0.147	&	0.165	&	16	&	\\
			4416.97	&	4417.59	&	\ion{O}{II}		&	V5	&	0.121	&	0.135	&	18	&	\\
			4416.97	&			&	\ion{O}{II}			&	V5		&			&			&		&	\\
			4428.54	&	4429.16	&	\ion{Ne}{II}		&	V57		&	0.065	&	0.073	&	25	&	\\
			4430.94	&	4431.56	&	\ion{Ne}{II} 		&	V61 		&	0.042	&	0.047	&	30	&	\\
			4432.74	&	4433.34	&	\ion{N}{II}			&	V55a	&	0.076	&	0.085	&	23	&	\\
			4434.60	&	4435.21	&	\ion{O}{III}	&	G[9/2]--H[11/2]o	&	0.040	&	0.045	&	31	&	\\
			4437.55	&	4438.16	&	\ion{He}{I}			&	1P$_{0}$-1S	&	0.065	&	0.072	&	25	&	\\
			4439.46	&	4439.95	&	\ion{Ar}{II}			&	2D$_{0}$-2D	&	0.022	&	0.025	&	42	&	\\
			4439.88	&			&	\ion{Ar}{II}			&	2D$_{0}$-2D	&			&			&		&	\\
			4457.05	&	4457.70	&	\ion{Ne}{II}		&	V61a	&	0.038	&	0.043	&	32	&	\\
			4457.26	&			&	\ion{Ne}{II}		&	V61d	&			&			&		&	\\
			4466.43	&	4467.04	&	\ion{O}{II}	        &	V86b	&	0.044	&	0.049	&	30 &	\\
			4469.45	&	4470.06	&	\ion{O}{II}	        &	4D -- 4P	&	0.032	&	0.035	&	35	&	\\
			4471.47	&	4472.12	&	\ion{He}{I}			&   V14	        &	6.251	&	6.897	&	3	&	\\
			4477.90	&	4478.17	&	\ion{O}{II}	        &	V88	&	0.115	&	0.127	&	19	&	\\
			4481.21	&	4481.85	&	\ion{Mg}{II}		&	V4			&	0.039	&	0.042	&	32	&	\\
			4489.45	&	4490.07	&	\ion{O}{II}			&	V86b    	&	0.022	&	0.024	&	43	& \\
			4491.23	&	4491.86	&	\ion{O}{II}			&	V86a		&	0.097	&	0.107	&	20	&	\\
			4498.92	&	4499.62	&	\ion{Ne}{II}		&	V64c     	&	0.025	&	0.028	&	39	& \\
			4499.12	&			&	\ion{Ne}{II}		&	V64c 	&			&			&		&	\\
			4510.91	&	4511.53	&	\ion{N}{III}		&	V3	&	0.221	&	0.241	&	13	&	\\
			4514.86	&	4515.48	&	\ion{N}{III}	    &	V3	&	0.042	&	0.046	&	31	&	\\
			4518.15	&	4518.75	&	\ion{N}{III}	    &	V3	&	0.074	&	0.081	&	23	&	\\
			4520.69	&	4521.34	&	\ion{Si}{II}		&	2D-2P$_{0}$	&	0.056	&	0.061	&	27	&	\\
			4523.58	&	4524.22	&	\ion{N}{III}	    &	V3	&	0.051	&	0.056	&	28	&	\\
			4530.41	&	4531.04	&	\ion{N}{II}			&	V58b	&	0.104	&	0.113	&	19	&	\\
			4530.86	&			&	\ion{N}{III}		&	V3	&			&			&		&	\\
			4534.58	&	4535.15	&	\ion{N}{III}	    &	V3	&	0.037	&	0.040	&	33	&	\\
			4541.59	&	4542.24	&	\ion{He}{II}	    &	4.9	&	0.074	&	0.080	&	23	&	\\
			4544.85	&	4545.47	&	\ion{N}{III}	    &	V12	&	0.020	&	0.022	&	44	&	\\
			4552.52	&	4553.22	&	\ion{N}{II}			&	V58a	&	0.068	&	0.074	&	24	&	\\
			4562.60	&	4563.22	&	\ion{Mg}{I}]	    &		&	0.072	&	0.078	&	23	&	\\
			4571.10	&	4571.73	&	\ion{Mg}{I}]	&		&	0.499	&	0.536	&	9	&	\\
			4590.97	&	4591.60	&	\ion{O}{II}		&	V15	&	0.116	&	0.124	&	18	&	\\
			4595.95	&	4596.79	&	\ion{O}{II}		&	V15	&	0.086		& 0.092			&	21	&	\\
			4596.18	&			&	\ion{O}{II}		&	V15	&		&		&		&	\\
			4602.13	&	4602.69	&	\ion{O}{II}	&	V92b	&	0.118	&	0.126	&	18	&	\\
			4607.16	&	4607.81	&	\ion{N}{II}	&	V5	&	0.033	&	0.035	&	35	&	\\
			4609.44	&	4610.07	&	\ion{O}{II}		&	V92a	&	0.236	&	0.251	&	13	&	\\
			4610.20	&	4610.87	&	\ion{O}{II}		&	V92c 	&	0.141	&	0.150	&	17	&	\\
			4613.68	&	4614.43	&	\ion{O}{II}	&	V92b	&	0.037	&	0.039	&	33	&	\\
			4620.26	&	4621.08	&	\ion{C}{II}	&		&	0.057	&	0.060	&	26	&  ?	\\
			4621.39	&	4622.02	&	\ion{N}{II}			&	V5		&	0.056	&	0.060	&	26	&	\\
			4630.54	&	4631.18	&	\ion{N}{II}			&	V5		&	0.281	&	0.298	&	12	&	\\
			4634.14	&	4634.78	&	\ion{N}{III}			&	V2	&	1.301	&	1.375	&	6 		& \\
			4638.86	&	4639.50	&	\ion{O}{II}			&	V1	&	0.645	&	0.681	&	8	&	\\
			4640.64	&	4641.28	&	\ion{N}{III}			&	V2	&	1.820	&	1.920	&	5	&	\\
			4641.83	&	4642.47	&	\ion{O}{II}			&	V1	&	1.302	&	1.373	&	6	&	\\
			4643.06	&	4643.73	&	\ion{N}{II}			&	V5	&	0.047	&	0.050	&	29	&	\\
			4647.45	&	4648.09	&	\ion{C}{III} 	    &	V1	&	0.108	&	0.113	&	19  & \\
			4649.13	&	4649.78	&	\ion{O}{II}			&	V1	&	1.353	&	1.424	&	6	&	\\
			4650.84	&	4651.49	&	\ion{O}{II} 		&	V1	&	0.404	&	0.425	&	10	&	\\
			4656.39	&	4657.02	&	\ion{Ne}{I}			&	3/2[1/2]-1/2[1/2]$_{0}$	&	0.077&0.081&23 & \\ 
			4658.05	&	4658.67	&	[\ion{Fe}{III}] 	&	3F 	&	0.057	&	0.060	&	26	& 	\\
			4661.63	&	4662.27	&	\ion{O}{II}			&	V1	&	0.479	&	0.503	&	9	&	\\
			4669.27	&	4669.98	&	\ion{O}{II} 	    &	V89b &	0.023	&	0.024	&	42	&	\\
			4673.73	&	4674.38	&	\ion{O}{II}			&	V1	&	0.063	&	0.066	&	25	&	\\
			4676.24	&	4676.87	&	\ion{O}{II}			&	V1	&	0.307	&	0.321	&	11	&	\\
			4678.11	&	4678.76	&	\ion{N}{II} 	    &	V61b	&	0.019	&	0.020	&	45	&	\\
			4685.71	&	4686.40	&	\ion{He}{II}		&	4-3	&	3.952	&	4.122	&	3	&	?	\\
			4694.66	&	4695.31	&	\ion{N}{II}	        &	V61a	&	0.032	&	0.033	&	35	&	\\
			4696.35	&	4696.98	&	\ion{O}{II}			&	V1	&	0.051	&	0.053	&	28	&	\\
		\end{tabular}
	\end{table*}
	
	\begin{table*}
		\contcaption{}
		\label{tab:continued4}
		\begin{tabular}{c c c c c c c c}
			\hline
			$\lambda_{0}$ & $\lambda_{\text{obs}}$ & Ion & Identification & F & I & Error
			(\%) & Notes \\
			\hline
			\hline
			4699.22	&	4699.83	&	\ion{O}{II}			&	V25	&	0.030	&	0.031	&	36	&	\\
			4703.16	&	4704.02	&	\ion{O}{II}			&	2D$_{0}$-2F 		&	0.055	&	0.057	&	27	&	\\
			4703.36	&			&	 \ion{Ar}{II}		&			&			&			&		&	\\
			4705.35	&	4705.99	&	\ion{O}{II}			&	V25 		&	0.039	&	0.040	&	32	&	\\
			4710.00	&	4710.67	&	\ion{O}{II}			&		&	0.035	&	0.037	&	33	&	\\
			4711.37	&	4712.03	&	[\ion{Ar}{IV}]		&	1F	&	1.291	&	1.338	&	6	&	\\
			4713.17	&	4713.83	&	\ion{He}{I}			&	V12	&	0.690	&	0.714	&	8	&	\\
			4740.23	&	4740.88	&	[\ion{Ar}{IV}]		&	1F	&	1.453	&	1.495	&	5	&	\\
			4751.34	&	4752.02	&	\ion{O}{II}			&		&	0.075	&	0.077	&	23	&	\\
			4762.31	&	4762.89	&	\ion{C}{I}			&	6	&	0.027	&	0.028	&	38	&	\\
			4788.13	&	4788.82	&	\ion{N}{II} 			&	20	&	0.020	&	0.021	&	44	&	\\
			4802.23	&	4803.11	&	\ion{C}{II}	&		&	0.068	&	0.069	&	24	& \\
			4803.27	&	4803.93	&	\ion{N}{II}	&	V20&	0.073	&	0.074	&	23	&	\\
			4815.51	&	4816.20	&	\ion{S}{II}		&	9		&	0.025	&	0.026	&	39	&	\\
            4859.36	&	4860.03	&	\ion{He}{ii}	&	8-4 &	0.181	&	0.181	&	15	&	\\
			4861.33	&	4862.00	&	\ion{H}{I}		&	H$\beta$	&	100.000	&	99.986	&	2	&	\\
			4890.86	&	4891.53	&	\ion{O}{II}	&	V28	&	0.028	&	0.028	&37	&	\\
			4901.28	&	4902.13	&	\ion{S}{II}	&		&	0.114	&	0.113	&	19	&	\\
			4902.65	&	4903.32	&	\ion{Si}{II}	&		&	0.026	&	0.026	&	39	&	\\
			4921.93	&	4922.62	&	\ion{He}{I}	&	V48&	1.701	&	1.678	&	5	&	\\
			4924.53	&	4925.20	&	\ion{O}{II}	&	V28	&	0.152	&	0.150	&	16	&	\\
			4931.32	&	4931.91	&	[\ion{O}{III}]	&	F1	&	0.091	&	0.090	&	21	&	\\
			4958.91	&	4959.63	&	[\ion{O}{III}]	&	1	&	269.852	&	264.011	&	2	&	\\
			5006.84	&	5007.56	&	[\ion{O}{III}]	&	F1		&	746.310	&	722.761	&	2	&	Saturated\\
			5015.70	&	5016.39	&	\ion{He}{I}		&	V4		&	2.819	&	2.725	&	4	&	\\
			5025.66	&	5026.32	&	\ion{N}{II}		&	1V9		&	0.018	&	0.017	&	43	&	\\
			5032.13	&	5032.94	&	\ion{C}{II}		&	2P-2D	&	0.039	&	0.038	&	29	&	\\
			5041.12 &	5041.74	&	\ion{Si}{II}		&	V5		&	0.329	&	0.316	&	10	&	\\
			5055.98	&	5056.86	&	\ion{Si}{II}		&	V5		&	0.131	&	0.125	&	16	&	\\
			5121.82	&	5122.57	&	\ion{C}{II}		&	V12		&	0.048	&	0.045	&	26	&	\\
			5191.82	&	5192.43	&	[\ion{Ar}{III}]	&	F3	&	0.055	&	0.052	&	24	&	\\
			5197.90	&	5198.63	&	[\ion{N}{I}]	&	F1	&	0.715	&	0.667	&	7	&	\\
			5200.06	&	5200.97	&	[\ion{Fe}{II}]	&	F19	&	0.512	& 	0.478	&		& \\
			5200.26	&			&	[\ion{N}{I}]	&	F1	&		&			&	8		&	\\
			5270.40	&	5271.19	&	[\ion{Fe}{III}]	&	F1	&	0.032	&	0.030	&	32	&	\\
			5342.38	&	5343.15	&	\ion{C}{II}		&	17.06	&	0.127	&	0.116	&	16	&	\\
			5346.02	&	5346.62	&	[\ion{Kr}{IV}]	&	1F	&	0.019	&	0.017	&	42	&	? \\
			5405.15	&	5405.91	&	\ion{Ne}{II}	&	2[5]$_{0}$-2[6]	&	0.031	&	0.028	&	33	&	\\
			5411.52	&	5412.30	&	\ion{He}{II}	&	4.7	&	0.334	&	0.301	&	10	&	\\
			5453.81	&	5454.67	&	\ion{S}{II}	&	V6	&	0.030	&	0.027	&	33	&	\\
			5495.67	&	5496.45	&	\ion{N}{II}	&	V29	&	0.018	&	0.016	&	43	&	\\
			5517.71	&	5518.45	&	[\ion{Cl}{III}] 	&	F1	&	0.666	&	0.590	&	7	&	\\
			5537.88	&	5538.62	&	[\ion{Cl}{III}] 	&	F1	&	0.898	&	0.793	&	6	&	\\
			5577.34	&	5578.46	&	[\ion{O}{I}]	& F3		&	0.044	&	0.039	&	27	&	\\
			5666.64	&	5667.41	&	\ion{N}{II}		&	V3	&	0.287	&	0.249	&	11	&	\\
			5676.02	&	5676.80	&	\ion{N}{II}		&	V3	&	0.121	&	0.105	&	17	&	\\
			5679.56	&	5680.35	&	\ion{N}{II}		&	V3	&	0.674	&	0.584	&	7	&	\\
			5686.21	&	5687.01	&	\ion{N}{II}		&	V3	&	0.081	&	0.070	&	20	&	\\
			5710.76	&	5711.54	&	\ion{N}{II}		&	V3	&	0.121	&	0.104	&	17	&	\\
			5739.73	&	5740.61	&	\ion{Si}{III}  	&	V4 	&	0.018	&	0.016	&	43	&	\\
			5747.33	&	5748.22	&	\ion{O}{II}		&	2D$_{0}$-2F	&	0.057	&	0.049	&	24	&	\\
			5754.64	&	5755.43	&	[\ion{N}{II}]	&	3F	&	5.720	&	4.907	&	3	&	\\
			5867.74	&	5868.56	&	[\ion{Kr}{IV}] 		&	4S-2D	&	0.024	&	0.021	&	22	& \\
			5875.60	&	5876.48	&	\ion{He}{I}			&	V11	&	22.009	&	18.602	&	2	& \\
			5927.78	&	5928.59	&	\ion{N}{II} &	V28 	&	0.020	&	0.016&	25	& \\
			5931.85	&	5932.72	&	\ion{He}{II}	&	5.25	&	0.066	&	0.055	&	14	& ?\\
			5940.24	&	5941.07	&	\ion{N}{II}	&	V28	&	0.013	&	0.011	&	31	& \\ 
			5941.65	&	5942.46	&	\ion{N}{II}	&	V28	&	0.092	&	0.077	&	12	& \\
			5952.58	&	5953.40	&	\ion{He}{II}	&	5.23	&	0.026	&	0.022	&	22	& ?\\
			5957.56	&	5958.24	&	\ion{Si}{II}		&	V4	&	0.010	&	0.008	&	36	& \\
			5978.98	&	5979.81	&	\ion{Si}{II}	&	V4	&	0.025	&	0.021	&	22	& \\
			6074.13&	6074.97	&	\ion{He}{II}	&	5.20	&	0.015	&	0.013	&	28	& \\
			6101.83	&	6102.62	&	[\ion{K}{IV}]	&	F1	&	0.090	&	0.074	&	12	& \\
		\end{tabular}
	\end{table*}

	\begin{table*}
		\contcaption{}
		\label{tab:continued5}
		\begin{tabular}{c c c c c c c c}
			\hline
			$\lambda_{0}$ & $\lambda_{\text{obs}}$ & Ion & Identification & F & I & Error
			(\%) & Notes \\
			\hline
			\hline
			6151.43	&	6152.23	&	\ion{C}{II}		&	16.04	&	0.087	&	0.072	&	12	& \\
			6157.42	&	6158.40	&	\ion{Ni}{II}	&		&	0.023	&	0.019	&	23	& ? \\
			6233.82	&	6234.43	&	\ion{He}{II}	&	5.17	&	0.019	&	0.016	&	25	& \\
			6300.30	&	6301.19		&	[\ion{O}{I}	]	&	F1	&	8.701	&	7.066	&	2	& \\
			6312.10	&	6312.97		&	[\ion{S}{III}]	&	 3F	&	1.830	&	1.484	&	3	& \\
			6347.11	&	6348.01		&	\ion{Si}{II}	&	V2	&	0.158	&	0.126	&	9	& \\ 
			6363.78	&	6364.68		&	[\ion{O}{I}]	&	F1	&	2.944	&	2.375	&	3	& \\
			6371.36	&	6372.26		&	\ion{Si}{II}	&	V2	&	0.236	&	0.189	&	7	& \\
			6402.25	&	6403.10		&	\ion{Ne}{I}		&	V1	&	0.014	&	0.011	&	30	& ?\\
			6406.38	&	6407.24		&	\ion{He}{II}	&	5-15	&	0.013	&	0.010	&	31	& \\
			6461.95	&	6462.75		&	\ion{C}{II}		&	17.04	&	0.315	&	0.250	&	6	& \\
			6482.05	&	6482.96		&	\ion{N}{II}		&	V8	&	0.030	&	0.023	&	20	& \\
			6485.30	&	6485.907	&	[\ion{Fe}{II}]	&	b4P-a2S	&	0.028	&	0.022	&	21	& \\
			6486.46	&	6487.33		&	\ion{O}{II}		&	G[3]$_{0}$-1[4]	&	0.029	&	0.023	&	21	&\\
     		6527.24	&	6528.053		&	[\ion{N}{II}] &	F1		&	0.041	&	0.032	&	17	& \\ 
			6548.03	&	6549.003		&	[\ion{N}{II}] &	F1	&	59.658	&	46.867	&	2	& \\
			6560.18	&	6561.08		&	\ion{He}{II}	&	6-4	&	0.763	&	0.598	&	4	& \\
			6562.82	&	6563.72		&	\ion{H}{I}	&	H$\alpha$		&	379.784	&	297.894	&	2	& \\
			6578.05	&	6578.96		&	\ion{C}{II}		&	V2	&	0.896	&	0.701	&	4	& \\ 
			6583.41	&	6584.38		&	[\ion{N}{II}]	&	F1	&	185.048	&	144.835	&	2	& \\
			6678.15	&	6679.08		&	\ion{He}{I}	&	V46	&	7.528	&	5.834	&	2	& \\
			6683.27	&	6684.19		&	\ion{Ne}{II}	&	2[2]$_{0}$-4D	&	0.026	&	0.020	&	22	& \\
			6716.47	&	6717.40		&	[\ion{S}{II}]  	&	F2	&	12.618	&	9.740	&	2	& \\
			6721.39	&	6722.28		&	\ion{O}{II}	&	V4	&	0.012	&	0.009	&	32	& \\
			6730.85	&	6731.79		&	[\ion{S}{II}]  		&	F2	&	20.368	&	15.699	&	2	& \\
			6744.39	&	6745.21		&	\ion{N}{II}	&	3.000	&	0.008	&	0.006	&	38	& \\
			6780.06	&	6780.97			&	\ion{C}{II}	&	V14		&	0.017	&	0.013	&	27	& \\
			6809.99	&	6811.11	&	\ion{N}{II}		&	V54	&	0.005	&	0.004	&	38	&	\\
			6821.16	&	6822.31	&	[\ion{Mn}{III}]	&		&	0.008	&	0.006	&	32	&	?\\
			6933.89	&	6934.88	&	\ion{He}{I}		&	1/13	&	0.014	&	0.011	&	26	&	\\
			6989.47	&	6990.48	&	\ion{He}{I}		&	1/12	&	0.021	&	0.016	&	22	&	\\
	      	7062.26	&	7063.35	&	\ion{He}{I}		&	1/11	&	0.037	&	0.028	&	16	&	\\
			7065.28	&	7066.25	&	\ion{He}{I}		&	V10	&	9.924	&	7.458	&	2	&	\\
			7135.78	&	7136.79	&	[\ion{Ar}{III}]	&	F1	&	32.270	&	24.076	&	2	&	\\
			7160.61	&	7161.55	&	\ion{He}{I}		&	1/10			&	0.055	&	0.041	&	14	&	\\
			7231.34	&	7232.32	&	\ion{C}{II}		&	V3		&	0.660	&	0.487	&	4	&	\\
			7236.42	&	7237.46	&	\ion{C}{II}		&	V3	&	1.529	&	1.129	&	3	&	\\
			7262.76	&	7263.95	&	[\ion{Ar}{IV}]	&	F2	&	0.037	&	0.027	&	16	&	\\
			7281.35	&	7282.39	&	\ion{He}{I}		&	V45		&	1.175	&	0.864	&	3	&	\\
			7318.92	&	7320.95	&	[\ion{O}{II}]	&	F2		&	6.587		& 4.822	& 2	&	\\
			7319.99	&			&	[\ion{O}{II}]	&	F2	&		&		&		&	\\
			7329.66	&	7331.23	&	[\ion{O}{II}]	&	F2	&	5.367	&	3.925	&	2	&	\\
			7330.73	&			&	[\ion{O}{II}]	&	F2	&			&			&		&	\\
			7499.85	&	7500.85	&	\ion{He}{I}		&	1/8	&	0.087	&	0.062	&	11	&	\\
			7519.49	&	7520.78	&	\ion{C}{II}	&	16.08	&	0.009	&	0.007	&	32	&	\\
			7519.86	&			&	\ion{C}{II}	&	16.08	&			&			&		&	\\
			7530.54	&	7531.51	&	[\ion{Cl}{IV}]	&	F1	&	0.377	&	0.270	&	5	&	\\
			7535.40	&	7536.07	&	[\ion{Xe}{IV}]	&	4S-2D	&	0.008	&	0.006	&	35	&	\\
			7751.10	&	7752.19	&	[\ion{Ar}{III}]	&	F2	&	8.011	&	5.615	&	2	&	\\
			7816.13	&	7817.21	&	\ion{He}{I}		&	1/7	&	0.112	&	0.078	&	10	&	\\
			8045.63	&	8046.85	&	[\ion{Cl}{IV}]	&	F1	&	0.864	&	0.588	&	4	&	\\
			8233.21	&	8234.33	&	\ion{H}{I}	&	P 50	&	0.049	&	0.033	&	14	&	\\
			8234.43	&	8235.55	&	\ion{H}{I}	&	P 49    &	0.062	&	0.042	&	13	&	\\
			8235.74	&	8237.93	&	\ion{H}{I}		&	P 48	&	0.438		&	0.293			&	5	&	\\	
			8236.79	&			&	\ion{He}{II}	&	5.9	&		&		&		&	\\
			8237.13	&			&	\ion{H}{I}		&	P 47	&			&			&		&	\\
			8238.61	&	8239.73	&	\ion{H}{I}	&	P 46	&	0.081	&	0.054	&	11	&	\\
			8240.19	&	8241.33	&	\ion{H}{I}	&	P 45	&	0.072	&	0.048	&	12	&	\\
			8241.89	&	8242.91	&	\ion{H}{I}	&	P 44	&	0.063	&	0.042	&	13	&	\\
			8243.70	&	8244.87	&	\ion{H}{I}	&	P 43	&	0.082	&	0.055	&	11	&	\\
			8245.64	&	8246.79	&	\ion{H}{I}	&	P 42	&	0.082	&	0.055	&	11	&	\\
			8247.72	&	8248.87	&	\ion{H}{I}	&	P 41	&	0.087	&	0.058	&	11	&	\\
			8249.97	&	8251.10	&	\ion{H}{I}	&	P40	&	0.082	&	0.055	&	10	&	\\
		\end{tabular}
	\end{table*}
	
		\begin{table*}
		\contcaption{}
		\label{tab:continued6}
		\begin{tabular}{c c c c c c c c}
			\hline
			$\lambda_{0}$ & $\lambda_{\text{obs}}$ & Ion & Identification & F & I & Error
			(\%) & Notes \\
			\hline
			\hline
			8252.40	&	8253.60	&	\ion{H}{I}	&	P 39	&	0.094	&	0.063	&	9	&	\\
			8255.02	&	8256.01	&	\ion{H}{I}	&	P 38	&	0.076	&	0.051	&	10	&	\\
			8257.86	&	8258.99	&	\ion{H}{I}	&	P 37	&	0.096	&	0.064	&	10	&	\\
			8260.94	&	8262.13	&	\ion{H}{I}	&	P 36	&	0.112	&	0.075	&	8	&	\\
			8264.28	&	8265.54	&	\ion{H}{I}	&	P 35	&	0.148	&	0.099	&	8	&	\\
			8267.94	&	8269.10	&	\ion{H}{I}	&	P 34	&	0.122	&	0.082	&	8	&	\\
			8271.93	&	8273.09	&	\ion{H}{I}	&	P 33	&	0.128	&	0.086	&	9	&	\\
			8276.31	&	8277.48	&	\ion{H}{I}	&	P 32    &	0.122	&	0.082	&	8	&	\\
			8281.12	&	8282.54	&	\ion{H}{I}	&	P 31	&	0.109	&	0.073	&	8	&	\\
			8286.43	&	8287.41	&	\ion{H}{I}	&	P 30	&	0.124	&	0.082	&	8	&	\\
			8292.31	&	8293.41	&	\ion{H}{I}	&	P 29	&	0.172	&	0.114	&	8	&	\\
			8298.83	&	8299.93	&	\ion{H}{I}	&	P 28	&	0.183	&	0.122	&	7	&	\\
			8306.11	&	8307.28	&	\ion{H}{I}	&	P 27	&	0.225	&	0.150	&	6	&	\\
			8314.26	&	8315.43	&	\ion{H}{I}	&	P26	&	0.226	&	0.150	&	6	&	\\
			8323.42	&	8324.57	&	\ion{H}{I}	&	P 25	&	0.250	&	0.166	&	6	&	\\
			8333.78	&	8334.89	&	\ion{H}{I}	&	P 24	&	0.290	&	0.192	&	6	&	\\
			8345.55	&	8346.70	&	\ion{H}{I} 	&	P23	&	0.308	&	0.204	&	5	&	\\
			8359.00	&	8360.20	&	\ion{H}{I}	&	P22	&	0.365	&	0.242	&	5	&	\\
			8361.71	&	8362.92	&	\ion{He}{I}	&	V68	&	0.185	&	0.122	&	7	&	\\
			8374.48	&	8375.72	&	\ion{H}{I}	&	P 21	&	0.372	&	0.246	&	5	&	\\
			8392.40	&	8393.62	&	\ion{H}{I}	&	P 20	&	0.430	&	0.285	&	4	&	\\
			8413.32	&	8414.72	&	\ion{H}{I}	&	P 19	&	0.680	&	0.449	&	4	&	\\
			8421.95 &	8423.10	&	\ion{He}{I}	&	6/18	&	0.030	&	0.020	&	16	&	\\
			8433.85	&	8434.38	&	[\ion{Cl}{III}] 	&	3F	&	0.007	&	0.005	&	34	&	?\\
			8437.96	&	8439.14	&	\ion{H}{I}	&	P 18	&	0.583	&	0.384	&	4	&	\\
			8446.25	&	8447.02	&	\ion{O}{I}	&	4 	&	0.095	&	0.062	&	10	&	\\
			8446.36	&			&	\ion{O}{I}	&	4	&			&			&		&	\\
			8446.76	&			&	\ion{O}{I}	&	4	&			&			&		&	\\
			8451.16	&	8452.23	&	\ion{He}{I} 	&	6/17	&	0.172	&	0.113	&	8	&	\\
			8453.61	&			&	\ion{He}{I}		& 7/17	&			&			&		&	\\
			8467.25	&	8468.45	&	\ion{H}{I}	&	P 17	&	0.710	&	0.467	&	4	&	\\
			8486.27	&	8487.54	&	\ion{He}{I}	&	6/16	&	0.031	&	0.020	&	16	&	\\
			8499.70	&	8500.99	&	[\ion{Cl}{III}]	&	3F	&	0.008	&	0.005	&	32	&	\\
			8502.48	&	8503.74	&	\ion{H}{I}	&	P16	&	0.777	&	0.509	&	4	&	\\
			8665.02	&	8666.03	&	\ion{H}{I} 	&	P 13	&	1.831	&	1.186	&	5	&	\\
			8680.28	&	8681.48	&	\ion{N}{I}	&	1	&	0.064	&	0.042	&	24	&	\\
			8727.13	&	8728.35	&	[\ion{C}{I}]	&	3F	&	0.116	&	0.075	&	20	&	\\
			8733.43	&	8734.68	&	\ion{He}{I}	&	6/12	&	0.088	&	0.057	&	20	&	\\
			8736.04	&	8737.31	&	\ion{He}{I}	&	7/12	&	0.032	&	0.021	&	34	&	\\
			8747.15	&	8748.35	&	\ion{He}{II}	&	6--24	&	0.040	&	0.026	&	34	&	\\
			8750.47	&	8751.68	&	\ion{H}{I}	&	P 12	&	1.931	&	1.244	&	5	&	\\
			8816.50	&	8817.83	&	\ion{He}{I}	&	10/12	&	0.018	&	0.011	&	46	&	\\
			8829.40	&	8831.00	&	[\ion{S}{III}]	&	3F	&	0.042	&	0.027	&	30	&	\\
			8845.38	&	8846.68	&	\ion{He}{I}		&	6/11	&	0.127	&	0.081	&	18	&	\\
			8862.79	&	8864.02	&	\ion{H}{I}	&	P 11	&	2.402	&	1.536	&	4	&	\\
			8996.99	&	8998.15	&	\ion{He}{I}	&	6/10	&	0.173	&	0.110	&	16	&	\\
			9014.91	&	9015.91	&	\ion{H}{I}	&	P10	&	2.608	&	1.654	&	4	&	\\
			9063.29	&	9064.62	&	\ion{He}{I}	&	4/8	&	0.196	&	0.124	&	14	&	\\
			9068.60	&	9070.25	&	[\ion{S}{III}]	&	F1	&	72.752	&	46.011	&	2	&	\\
			9123.60	&	9124.98	&	[\ion{Cl}{II}]	&	F1	&	0.150	&	0.095	&	16	&	\\
			9210.28	&	9211.71	&	\ion{He}{I}	&	6/9	&	0.236	&	0.148	&	14	&	\\
			9213.20	&	9214.51	&	\ion{He}{I}	&	7/9	&	0.056	&	0.035	&	26	&	\\
			9229.01	&	9230.59	&	\ion{H}{I}	&	P 9	&	4.930	&	3.094	&	3	&	\\
			9516.57	&	9518.13	&	\ion{He}{I}	&	4/7	&	0.136	&	0.084	&	18	&	\\
			9526.16	&	9527.65	&	\ion{He}{I}	&	6 / 8	&	0.328	&	0.203	&	12	&	\\
			9531.21	&	9532.52	&	[\ion{S}{III}]	&	F1	&	160.117	&	99.117	&	2	&	\\
			9545.97	&	9547.52	&	\ion{H}{I}	&	P 8	&	5.845	&	3.616	&	3	&	\\
			9824.13	&	9825.27	&	[\ion{C}{I}]	&	F1	&	0.226	&	0.138	&	14	&	\\
			9850.26	&	9851.63	&	[\ion{C}{I}]	&	F1	&	0.747	&	0.457	&	8	&	\\
			9982.46	&	9983.61	&	\ion{O}{II}	&	G[5]$_{0}$-2[6]	&	0.150	&	0.091	&	16	&	\\
			9988.54	&	9989.73	&	\ion{O}{II}	&	G[5]$_{0}$-2[6]	&	0.132	&	0.080	&	18	&	\\
			9990.08	&	9991.38	&	\ion{O}{II}	&	D[3]$_{0}$-0[4]	&	0.260	&	0.158	&	12	&	\\
			9991.48	&	9992.72	&	\ion{O}{II}	&	D[3]$_{0}$-0[4]	&	0.127	&	0.077	&	18	&	\\

		\end{tabular}
	\end{table*}
	
    	\begin{table*}
    	\label{tab:continued7}
		\contcaption{}
		\begin{tabular}{c c c c c c c c}
			\hline
			$\lambda_{0}$ & $\lambda_{\text{obs}}$ & Ion & Identification & F & I & Error
			(\%) & Notes \\
			\hline
			\hline
			10008.87&	10010.09&	\ion{Ne}{I}	&		&	0.131	&	0.080	&	18	&	\\
			10027.70&	10028.61&	\ion{He}{I}	&	6/7	&	0.390	&	0.237	&	10	&	\\ 
			10031.20&	10031.95&	\ion{He}{I}	&	7/7	&	0.135	&	0.082	&	18	&	\\
			10049.40&	10049.82&	\ion{H}{I}	&	P 7	&	8.842	&	5.360	&	3	&	\\
				\hline 
			\hline
		\end{tabular}
	\end{table*}

	\section*{Data availability}
	This paper includes data collected at the European Southern Observatory, Chile, programme ID 70.C-0008(A). The photoionization models presented in this paper are available under request to the authors.

	

	\bibliographystyle{mnras}
	\bibliography{manuscript_final.bib}

	


	\bsp	
	\label{lastpage}
\end{document}